\documentclass[11pt,a4paper]{article}
\pdfoutput=1 

\usepackage{jheppub}
\usepackage{amssymb,amsmath,amsthm,graphicx}
\usepackage{graphics, color}
\usepackage{subfigure}
\usepackage{latexsym}
\usepackage{bm}
\usepackage{epsfig}
\usepackage{textcomp}
\usepackage[utf8]{inputenc}
\allowdisplaybreaks[1]

\begin{document}

\title{Self Gravitating Spinning String Condensates}
\author{Jorge~E.~Santos and}
\author{Yoav~Zigdon}

\affiliation{DAMTP, Centre for Mathematical Sciences, University of Cambridge, Wilberforce Road, Cambridge CB3 0WA, UK}
\emailAdd{jss55@cam.ac.uk}
\emailAdd{yz910@cam.ac.uk}

\newcommand{\blue}{\color{blue}}
\newcommand{\red}{\color{red}}

\abstract{
In the context of the black hole/string transition, it is useful to produce Euclidean string backgrounds representing hot and self-gravitating strings. We utilise analytical and numerical methods to find a smooth, stationary rotating solution in the heterotic string theory at high temperatures. The solution describes a spinning winding-momentum condensate living in three non-compact dimensions, and its backreaction on the thermal cycle. At low temperatures, we expect a transition between our solution to an analytical continuation of an axionic Kerr black hole.}
\maketitle
\section{Introduction}
Black holes have been conjectured to transition to fundamental strings when their Hawking temperature approaches the Hagedorn temperature. Quantities like the Bekenstein-Hawking entropy of black holes can be extrapolated to a stringy regime where they agree with the corresponding quantities of highly-excited strings \cite{Smolin85},\cite{Susskind93}. Motivated by this idea, \cite{RussoSusskind94} computed the degeneracy of highly-excited rotating strings and compared the gyromagnetic ratios of heterotic black holes and strings. The ``correspondence principle'' describes the point where physical quantities of the two systems are expected to match and was demonstrated for a class of charged black strings and black branes in \cite{HP96}.

In a handful of special systems, one need not invoke an extrapolation of the entropy and temperature of strings and black holes to find a correspondence point. One example involves BPS strings that carry winding and momentum about a spatial circle and the BPS extremal black hole carrying the same charges in the heterotic string theory \cite{Sen95},\cite{Dabholkar04}. Other examples are two- and three-dimensional target spaces where exact CFT descriptions permit establishing the equality of entropies and temperatures at the correspondence point \cite{2AmitDavidEliezer}. In these examples, however, the energy is independent of the curvature of space. We will consider a system outside this special class. 

Horowitz and Polchinski (HP) \cite{HP97} found an Euclidean string background that describes a phase of hot and self-gravitating strings. This phase can be thought to exist between the ``free string phase'' at zero coupling and the ``black hole phase'' at relatively large coupling. The ``HP solution'' includes a winding condensate, representing a string mode that wraps around the thermal circle, and another mode that describes the size of the thermal circle in space. This solution admits a nonzero entropy at tree-level in the string coupling, one that scales with $\frac{1}{G_N}$ \cite{CM21}, a feature that was further investigated in \cite{RamyYoav1},\cite{RamyYoav2}.

The size of the solution in the non-compact dimensions is large in string units but smaller than the size of a free highly-excited string when the temperature deviates further away from the Hagedorn temperature; this allows for the possibility that the string further contracts as one decreases the temperature (or increases the string coupling), eventually reaching the size of a stringy black hole.  

More recently, \cite{CMW} generated charged versions of the Horowitz-Polchinski solution that include both a spatial circle and an Euclidean time circle. This reference also argued that in the context of tree-level Type II superstring theory, the black hole/string transition is not smooth due to distinct supersymmetric indices (or closely related protected quantities, depending on the number of dimensions) of the worldsheet CFTs describing string propagation in the corresponding Euclidean solutions. Versions of the HP solution in asymptotically Euclidean AdS space were found in \cite{Erez1},\cite{Erez2} \footnote{Some recent calculations of $\alpha'$ corrections to the Hagedorn temperature in asymptotically Euclidean AdS also appear in \cite{Minahan1},\cite{Minahan2},\cite{Harmark}. Reference \cite{Massai23} computed $\alpha'$ corrections to thermodynamic quantities of the heterotic two-charge black hole.}. Extensions of the solution to six or (slightly) more non-compact dimensions were constructed in \cite{David1},\cite{David2}, including a worldsheet description. The paper \cite{Jafferis23} studies HP-like solutions in asymptotically AdS$_3$ with NS-NS flux from a worldsheet perspective. The HP solution was embedded in string field theory \cite{Mazel:2024alu} and considered in the context of the Swampland program \cite{Bedroya:2024uva}.

A natural extension of previous works on the black hole/string transition is incorporating rotation. The paper \cite{Roberto} discusses aspects of this subject, for instance,  the transition of highly-spinning strings, with angular momentum that surpasses the ``Kerr bound'', to black objects. That paper pointed out likely evolution scenarios between the two phases, with intermediate steps that involve, depending on the number of spacetime dimensions, the emergence of black hole-string hybrids or black bars. 

In this paper, we construct Euclidean rotating HP solutions in the heterotic string theory. We have found smooth solutions only in the heterotic theory \cite{HeteroticString}; in the Type II theory, the winding condensate does not carry a momentum charge, and consequently, the leading order equation motion for the time-angle metric component, $G_{\tau \phi}$, implies that either $G_{\tau \phi}=0$ or admits an expression that leads to a curvature singularity at the center of space. By contrast, the heterotic string mode that becomes light near the Hagedorn temperature couples to the metric time-angle component, leading to the discovery of a smooth and rotating solution.  \footnote{This difference between the heterotic string theory and Type II string theory might be related to the difference pointed out in \cite{CMW}, where no obstruction to a smooth black hole/string transition in the heterotic theory was found, unlike the result in Type II.}

Another feature of our solution is that it includes NS-NS flux, in contrast to the original HP solution, where the flux vanishes. This difference is interpreted as the latter describing Lorentzian strings that perform random walks with no net rotation \cite{HP97}: Even though strings couple to the Kalb-Ramond field, when they perform random walks, the induced NS-NS fields from the various parts of the system cancel on average. In contrast, rotating strings produce a net $H_3$ field because no such cancellation occurs.

The original motivation of the work was to find a solution whose Lorentzian interpretation is that of rotating and hot, self-gravitating strings in four spacetime dimensions, to which the Kerr black hole could be connected as one varies the temperature or string coupling. Instead, we have found a solution with real angular momentum in Euclidean signature. When continuing to the Lorentzian signature, an imaginary angular momentum emerges. This is the main difference between our work and \cite{RussoSusskind94},\cite{Roberto}, which considered black holes and strings with real angular momentum in the Lorentzian signature. 

It is known that a double analytical continuation of the Kerr solution that involves Lorentzian to Euclidean time $t\to -i \tau$ and angular momentum parameter to an imaginary angular momentum parameter $ a\to i\alpha$ \cite{GibbonsHawking79} results in a solution sourced by Kaluza-Klein monopoles \cite{GrossPerry83},\cite{Dowker95}. Because our solution also admits a nonzero electric $H_3$ field, we expect that at stronger coupling it connects to a dyonic bound state charged under both winding and momentum. A Lorentzian interpretation is that the fluxes it includes give rise to the pair creation of strings with opposite charges, which can be viewed as a stringy version of the Schwinger effect.  

The paper runs as follows. In section \ref{Sec:EOM}, we derive equations of motion from an effective action that includes the standard bosonic term of supergravity and an explicit source term similar to \cite{HP97}. In section \ref{Sec:Sol}, we present numerical solutions and compute physical quantities like entropy and angular momentum. In section \ref{Sec:Conc}, we close with several conclusions. Appendices contain a derivation of the covariant derivative of the condensate in the heterotic string and the numerical convergence of our solutions.

\section{Equations of Motion}
\label{Sec:EOM}

In this section, we write equations of motion derived from an effective action of the heterotic string theory near the Hagedorn temperature for the light fields.    

The number of non-compact spatial dimensions is denoted by $d$, the number of these dimensions plus the periodic thermal direction is denoted by $D=d+1$, and the compact space we consider includes a  $T^{9-d}$ factor. The asymptotic space under consideration is 
\begin{align}
\label{AsymptoticSpace}
    R^d \times S^1 _{\beta} \times T^{9-d}. ~~~~
\end{align}
We will later set $d=3$, but at times, keep it general. 
The inverse Hagedorn temperature in this theory in a flat, constant dilaton background is \cite{HeteroticStringFree},\cite{AW}
 \begin{equation}
\label{InverseHagedorn}
	\beta_H  = (2+\sqrt{2})\pi \sqrt{\alpha'}.
\end{equation}
Unlike Type II superstring theory, the heterotic string theory with spontaneously broken supersymmetry admits string states with half-integer momentum numbers $n\in \frac{1}{2}\mathbb{Z}$. In order for fermionic modes to satisfy anti-periodic boundary conditions with respect to $S^1 _{\beta}$, the string wavefunction transforms as $(-1)^{n}$  as one cycles around this circle \cite{AW}.
The modes that become massless at the Hagedorn temperature in the heterotic string theory are labeled by the quantum numbers $|w=1,n=\frac{1}{2},N=0,\tilde{N}=0\rangle$ and $|w=-1,n=-\frac{1}{2},N=0,\tilde{N}=0\rangle$, where $w$ is the winding number about the circle, $n$ is the momentum number of the circle, $N,\tilde{N}$ are oscillator numbers of the left and right sectors of the theory. We denote these modes by $\chi$ and $\chi^*$, respectively. 
The mass squared of the two modes is
\begin{equation}
\label{MassSquared}
	m^2 = \frac{\beta^2}{4\pi^2 {\alpha^\prime}^2}+\frac{\pi^2}{\beta^2}-\frac{3}{\alpha'}. 
\end{equation}
Near the Hagedorn temperature, one can expand the above as
\begin{subequations}
\begin{equation}
m^2\approx \frac{\kappa_{het}}{\alpha^\prime}\left(\frac{\beta-\beta_H}{\beta_H}\right)+\mathcal{O}[(\beta-\beta_H)^2]\, ~,
\label{eq:m2}
\end{equation}
where
\begin{equation}
\kappa_{het}=4\sqrt{2}\,.
\end{equation}
\end{subequations}
The light fields are denoted by 
\begin{align}
\label{Fields}
 \{G_{\mu \nu},\sigma, A_{\mu}, B_{\tau \mu},B_{\mu \nu},\Phi_d, \chi,\chi^* \} ~,~ \mu,\nu=1,...,d,
\end{align} 
where lowercase Greek indices run over the non-compact dimensions. $G_{\mu \nu}$ are metric components in the string frame, $B_{\mu \nu}$ and $B_{\tau \mu}$ are NS-NS two-form components and $\Phi_d$ is the $d$-dimensional dilaton. The radion $\sigma$ controls the size of the circle and the metric $\tau-\mu$ component is related to the graviphoton $A_{\mu}$, which appears in the Kaluza-Klein reduction on the time circle
\begin{equation}
 {\rm d}s_D ^2 = {\rm d}s_{d} ^2 + e^{2\sigma} \left({\rm d}\tau + A_{\mu} {\rm d}x^{\mu} \right)^2. 
\end{equation}
The graviphoton $A_{\mu}$ has an associated field strength
\begin{equation}
    F_{\mu \nu} = \partial_{\mu} A_{\nu} - \partial_{\nu} A_{\mu}.
\end{equation}
The d-dimensional dilaton is defined in terms of the D-dimensional dilaton as 
\begin{equation}
    \Phi_d = \Phi_D - \frac{\sigma}{2}.
\end{equation}
The three-form NS-NS flux can be partitioned to components with a leg in the time direction, $H_{\tau \mu \nu}$, and 
\begin{equation}
\label{TildeH}
    \tilde{H}_{\mu \nu \lambda} = \partial_{\mu} B_{\nu \lambda} -A_{\mu} H_{\tau \nu \lambda} + \text{even~permutations}.
\end{equation}
For simplicity, we set the gauge fields arising as low-energy excitations of the heterotic string to zero.
The paper \cite{HP97} writes an effective action for winding modes near the Hagedorn temperature for the Type II and bosonic string theories. Reference \cite{SchulginTroost} does this for the heterotic theory by computing S-matrix elements with a circle of circumference approximately equal to $\beta_H$. 
Based on these references, the action describing the system  $I_{D}$ is a sum of two terms:
\begin{equation}
\label{Action}
	I_{D}=I_1+I_2,
\end{equation}
where
\begin{align}
    I_1 = \beta \int {\rm d}^d x \sqrt{G_d} e^{-2\Phi_d} \Big( D_{\mu} \chi D^{\mu} \chi^* + m^2 \chi \chi^* +\frac{\kappa_{het}}{\alpha'}\sigma \chi \chi^*\Big),
    \label{eq:I1N}
\end{align}
and
\begin{align}
\label{I_2}
    I_2=  -\frac{\beta}{2\kappa_0 ^2} \int {\rm d}^d x \sqrt{G_d} e^{-2\Phi_d} \Big(R_d +&4\partial_{\mu} \Phi_d \partial^{\mu} \Phi_d -\partial_{\mu} \sigma \partial^{\mu} \sigma - \frac{1}{4}e^{2\sigma} F_{\mu \nu} F^{\mu \nu}+ \nonumber\\
    &  -\frac{1}{4} e^{-2\sigma} H_{\tau \mu \nu} H_{\tau }^{~~\mu \nu}-\frac{1}{12} \tilde{H}_{\mu \nu \lambda} \tilde{H} ^{\mu \nu \lambda} \Big).
\end{align}
The string condensate $\chi$ couples to the graviphoton gauge field with a momentum charge unit $\frac{\pi}{\beta}$ and to the Kalb-Ramond two-form $B_{\tau \phi}$ with winding charge unit $\frac{\beta}{2\pi \alpha'}$. Thus, the covariant derivative takes the form \cite{SchulginTroost}
\begin{equation}
\label{CovariantDerivative}
	D_{\mu} \chi = \partial_{\mu} \chi + i \frac{\beta}{2\pi \alpha'}B_{\tau \mu}\chi +i\frac{\pi }{\beta}A_{\mu} \chi.
\end{equation}
The complex conjugate equation holds for $\chi^*$. We derive Eq.~(\ref{CovariantDerivative}) in appendix \ref{App:A} from a string amplitude calculation at tree-level.

The equations of motion derived from $I_{D}$ read
\begin{subequations}
	
	\begin{eqnarray}
		\label{chiEOMhet}
		&\frac{e^{2\Phi_d}}{\sqrt{G_d}}\partial_{\mu}\left(e^{-2\Phi_d}\sqrt{G_d}G^{\mu \nu} D_{\nu}\chi \right) =
		&\frac{\kappa_{het}}{ \alpha'}\sigma \chi+m^2 \chi-i\left( \frac{\beta}{2\pi \alpha'} B_{\tau}^{~ \mu} + \frac{\pi}{\beta} A^{\mu}\right)D_{\mu} \chi.\nonumber\\
		&
	\end{eqnarray}

	\begin{equation}
		\label{TwoForm2het}
		\partial_{\alpha} \left(\sqrt{G_d} e^{-2\Phi_d} \tilde{H}^{\alpha \mu  \nu} \right)=0.
	\end{equation}
	\begin{eqnarray}
		\label{C}
		&\frac{e^{2\Phi_d}}{ \sqrt{G_d}}\partial_{\alpha} \left(\sqrt{G_d} e^{-2\Phi_d} H^{\tau\alpha \mu}\right)=\frac{i \beta}{\pi \alpha'}\kappa_0^2\left(\chi D^{\mu} \chi^* - \chi^* D^{\mu} \chi \right)-\frac{1}{2}\tilde{H}^{\mu \alpha \lambda} F_{\alpha \lambda}.
	\end{eqnarray}
	
	\begin{eqnarray}
 \label{GraviphotonHet}
	&\frac{e^{2\Phi_d}}{\sqrt{G_d}}\partial_{\mu} \left( \sqrt{G_d} e^{-2\Phi_d+2\sigma} F^{\mu \nu}\right) = \frac{2\pi i \kappa_0^2 }{\beta}\left(\chi D^{\nu} \chi^* - \chi^* D^{\nu} \chi\right)-\frac{1}{2} \tilde{H}^{\nu \lambda \mu} H_{\tau\lambda \mu}.\nonumber\\
	&
	\end{eqnarray}
	
	\begin{eqnarray}
		\label{SigmaRhet}
		&\frac{e^{2\Phi_d}}{\sqrt{G_d}}\partial_{\mu} \left( e^{-2\Phi_d}\sqrt{G_d}\partial^{\mu}\sigma\right) =  \frac{1}{4}\left(  F_{\mu \nu}F^{\mu \nu}-H_{\tau \mu \nu}H^{~\mu \nu} _{\tau}\right)  + \frac{\kappa_{het}\kappa_0^2}{\alpha'}\chi \chi^*.
	\end{eqnarray}

\begin{eqnarray}
	\label{dilatonRhet}
	&R_d + 4 \nabla^2\Phi_d -4\partial^{\mu} \Phi_d \partial_{\mu} \Phi_d-\frac{1}{4}e^{2\sigma} F_{\mu \nu} F^{\mu \nu}-\partial_{\mu}\sigma \partial^{\mu}\sigma  -\frac{1}{4}H_{\tau \mu \nu}H^{\tau \mu \nu}-\frac{1}{12}\tilde{H}_{\mu \nu \lambda} \tilde{H} ^{\mu \nu \lambda}= \nonumber\\
	&= 2\kappa_0^2 \left[ D ^{\mu} \chi D_{\mu} \chi^* +m^2\chi \chi^*+\frac{\kappa_{het}}{\alpha'} \sigma\chi \chi^*\right].
\end{eqnarray}

\begin{eqnarray}
	\label{EinsteinRhet}
	&R_{d\mu \nu}-\partial_{\mu} \sigma \partial_{\nu} \sigma +2\nabla_{\mu}  \nabla_{\nu} \Phi_d -\frac{1}{2}H_{ \mu \lambda\tau}H_{\nu}^{~~ \lambda\tau}  -\frac{1}{4}\tilde{H}_{\mu \eta \lambda} \tilde{H}_{\nu } ^{~~\eta \lambda} -\frac{1}{2} e^{2\sigma} F_{\mu \alpha} F_{\nu} ^{~\alpha} =2\kappa_0 ^2 D_{\mu} \chi D_{\nu} \chi^*\,,\nonumber\\
	&
\end{eqnarray}
where we define the components of the current as
\begin{equation}
\label{Current}
	j_{\mu} \equiv i\left( \chi D_{\mu} \chi^* - \chi^* D_{\mu} \chi\right) = 2\chi \chi^* \left( \partial_{\mu} \text{arg}\,\chi + \frac{\beta}{2\pi \alpha'} B_{\tau \mu}+\frac{\pi}{\beta} A_{\mu}\right).
\end{equation}
\end{subequations}
The equations above are interpreted as the equations for the thermal scalar $\chi$ (which we will also refer to as the ``winding-momentum condensate''), the Kalb-Ramond field $B$, the graviphoton $A$, the radion $\sigma$ (valid for small $|\sigma|\ll 1$), the dilaton $\Phi_d$, and a linear combination of the spatial metric $G_{\mu \nu}$ and dilaton equations, respectively.

Now we prove that if $\tilde{H}_{\mu \nu \lambda}=0$ and there is a regular solution, then
\begin{equation}
\label{FvsH}
F_{\mu \nu} = e^{-4\sigma}\frac{2\pi^2 \alpha'}{\beta^2} H_{\tau \mu \nu}. 
\end{equation}
This follows from taking a linear combination of the graviphoton equation (\ref{GraviphotonHet}) and the NS-NS equation (\ref{C})
, that yields zero on the right-hand-side, and integrating over space. The constant of integration resulting from this must vanish in order to avoid a singularity of $F_{\mu \nu}-e^{-4\sigma}\frac{2\pi^2 \alpha'}{\beta^2} H_{\tau \mu \nu}$ at the origin. The ratio of momentum to winding charges is then given by Eq.~(\ref{FvsH}), as anticipated.

Next, we adopt a line element used for describing rotating, stationary boson stars \cite{Yoshida97}:
\begin{equation}
\label{LineElement}
	{\rm d}s^2 = a(r,\theta){\rm d}\tau^2 + b(r,\theta)\left( {\rm d}r^2 + r^2 {\rm d}\theta ^2\right)+r^2 \sin^2\theta \left({\rm d}\phi - \omega(r,\theta) {\rm d}\tau\right)^2.
\end{equation}
The angles $(\theta,\phi)$ parametrize a two-sphere with $0\leq \theta\leq \pi$ and $0\leq \phi<2\pi$, representing the usual polar and azimuthal angles on the two-sphere.

The line element (\ref{LineElement}) represents a configuration that rotates in a plane parametrized by the azimuthal angle $\phi$ with respect to the Euclidean time $\tau$. The angular velocity as a function of position is $\omega(r,\theta)$. 

Next, we introduce the following quantities
\begin{equation}
\label{Notations}
	\delta a(r,\theta) \equiv a(r,\theta)-1\quad\text{and}\quad 
	\delta b(r,\theta) \equiv b(r,\theta)-1.
\end{equation}
We approximate $|\delta a|\ll 1,|\delta b|\ll 1$ as well as $r^2 \omega^2 (r,\theta)\ll1 $. 
For the thermal scalar, we consider instead
\begin{equation}
\label{ansatz}
	\chi (r,\theta,\phi) = \tilde{\chi}(r,\theta) e^{im_{\phi}\phi}  ~,~ m_{\phi}\in \mathbb{Z}.
\end{equation}
Below we suppose that the only nonzero component of the $A$ field is in the azimuthal $\phi$ direction and depends only on $r$ and $\theta$. Similarly, we assume that the unique nonzero component of the $B$-field is $B_{\tau \phi} = B_{\tau \phi} (r,\theta)$.

We begin by looking at the $\chi$ equation (\ref{chiEOMhet}). The covariant derivative with respect to the azimuthal angle is
\begin{equation}
\label{Covariantphi}
	D_{\phi} \chi = im_{\phi}\chi +i\left(\frac{\beta}{2\pi \alpha'} B_{\tau\phi} + \frac{\pi}{\beta} A_{\phi}\right)\chi.
\end{equation}
Under the approximation $\Phi_d \approx \Phi_0$, $a\approx 1,\sqrt{G_d} \approx r^2 \sin\theta$, the $\chi$ equation can be written as
\begin{eqnarray}
	\label{chiEq2}
	&\partial^{\mu}D_{\mu}\chi \approx
	&\left[\frac{\beta ^2 (a(r,\theta)+r^2 \sin^2\theta \omega(r,\theta)^2)-\beta_H^2 }{4\pi^2 (\alpha')^2}+\frac{\pi^2}{\beta^2 a} \right] \chi-i\left( \frac{\beta}{2\pi \alpha'} B_{\tau}^{~ \phi} + \frac{\pi}{\beta} A^{\phi}\right)D_{\phi} \chi.\nonumber\\
	&
\end{eqnarray}
It follows that
\begin{align}
\label{chiSimplify}
	&\sum_{i\neq \phi}\partial^{i}\partial_{i}\chi  \approx  \left[\frac{\beta ^2 (a(r,\theta)+r^2 \sin^2\theta \omega(r,\theta)^2)-\beta_H^2 }{4\pi^2 (\alpha')^2}+\frac{\pi^2}{\beta^2 a} \right] \chi +\frac{1}{r^2 \sin^2\theta}\left(m_{\phi}+ \frac{\beta}{2\pi \alpha'} B_{\tau \phi} + \frac{\pi}{\beta} A_{\phi}\right)^2\chi.
\end{align}
Similarly to \cite{HP97} and \cite{CMW}, we want to assign scaling relations that allow one to simplify the equations
\begin{equation}\label{ScalingRelations}
	\chi (r,\theta,\phi) = \frac{\alpha'  m^2 \sqrt{4\pi}}{\kappa_{het}\kappa_0 \zeta} \hat{\chi} (r,\theta,\phi) ~,~ 	\delta a (r,\theta) = \frac{2\alpha' m^2}{\kappa_{het}\zeta}\hat{\delta a}(r,\theta) ~,~ r = \frac{\hat{r}}{m} \sqrt{\zeta}. 
\end{equation}
The eigenvalue parameter $\zeta$ is determined by numerics for each choice of $m_{\phi}$. Fig.~\ref{fig:zeta} in section \ref{Sec:Sol} presents a relevant graph of $\zeta$ plotted against $m_{\phi}$.\\
The scaling relations assigned to $A_{\phi}$ and $B_{\tau \phi}$ are determined as follows. First, we consider $\tilde{H}_{r\theta \phi}=0$ and $B_{\tau \theta}=0,B_{\tau r}=0$, which imply that the NS-NS flux threading the three spatial dimensions $(r,\theta,\phi)$:
\begin{equation}
    H_{r\theta \phi} = A_{\phi} H_{\tau r \theta}=0.
\end{equation}
Second,  Eq.~(\ref{C}),
\begin{align}
\label{BLaplacian1}
	&\frac{1}{r^2}\partial_r\left(   r^4 \sin^2\theta\,\partial_r B^{\tau \phi}\right)+ \frac{1}{ \sin \theta}\partial_{\theta} \left( \sin^3\theta \,\partial_{\theta} B^{\tau \phi}\right)\approx\frac{\beta \kappa_0 ^2}{\pi \alpha'} 2\chi \chi^*\left(m_{\phi} + \frac{\beta}{2\pi \alpha'}B_{\tau \phi} + \frac{\pi}{\beta}A_{\phi}\right).
\end{align}
This implies that
\begin{align}
\label{BLaplacian}
	&\frac{1}{r^2}\partial_r\left[   r^4 \partial_r \left(\frac{B_{\tau \phi}}{r^2}\right)\right]+ \frac{1}{r^2 \sin \theta}\partial_{\theta} \left[ \sin^3\theta\,\partial_{\theta} \left(\frac{B_{\tau \phi}}{\sin^2\theta}\right)\right]\approx\frac{\beta \kappa_0 ^2}{\pi \alpha'} 2\chi \chi^*\left(m_{\phi} + \frac{\beta}{2\pi \alpha'}B_{\tau \phi} + \frac{\pi}{\beta}A_{\phi}\right).
\end{align}
The left-hand-side scales like $\frac{m^2}{\zeta}$ times the scaling of the B-field while the right-hand-side has a term that scales like $\frac{m_{\phi}}{\zeta^2} \times {\alpha^\prime}^2 m^4$. One can verify a posteriori that the remaining terms in Eq.~(\ref{BLaplacian}) are subleading, justifying their neglect. Consequently, one can assign
\begin{equation}
\label{Bscaling}
	B_{\tau \phi } = \frac{8\beta m_{\phi} }{\kappa_{het}^2 }\frac{\alpha' m^2}{\zeta}\hat{B} _{\tau \phi}.
\end{equation} 
It is possible to add an integer times $\frac{2\pi \alpha'}{\beta}$ to expression (\ref{Bscaling}). Similarly, we assign
\begin{equation}
\label{Ascaling}
	A_{\phi} = \frac{(4\pi)^2   m_{\phi}}{\beta \kappa_{het}^2}\frac{ {\alpha^\prime}^2 m^2}{ \zeta }  \hat{A}_{\phi}.
\end{equation}
Since $A_{\phi} \approx -r^2 \sin^2\theta\,\omega(r,\theta)$, one has
\begin{equation}
	\label{OmegaScaling}
	\omega  = \frac{(4\pi)^2   m_{\phi} {\alpha^\prime}^2  m^4}{\beta \kappa_{het}^2\zeta^2} \hat{\omega}.
\end{equation}
Equipped with the scaling relations, one can write the equation (\ref{chiSimplify}) as
\begin{align}
\label{ChiEqfinal?}
	&\hat{\nabla}^2 \hat{\chi} \approx (\zeta + \delta \hat{a})\hat{\chi}.
\end{align}
The scaling relations also simplify considerably Eq.~(\ref{SigmaRhet}), where one should identify $\sigma = \frac{1}{2} \delta a$. The result is
\begin{equation}
\label{RadionEq}
	\nabla^2 \delta \hat{a} \approx 4\pi \hat{\chi} \hat{\chi}^*.
\end{equation}
The rescaled Kalb-Ramond field equation is given by
\begin{equation}
\label{BoxB}
\frac{1}{\hat{r}^2} \partial_{\hat{r}}  \left( \hat{r}^4 \sin^2\theta\, \partial_{\hat{r}}\hat{B}^{\tau \phi}\right)+\frac{1}{\sin\theta}\partial_{\theta} \left(\sin^3\theta\,\partial_{\theta} \hat{B}^{\tau \phi}\right)\approx \hat{\chi} \hat{\chi}^*.
\end{equation}
The rescaled graviphoton field equation is given by
\begin{equation}
	\label{BoxA}
	\frac{1}{\hat{r}^2} \partial_{\hat{r}}  \left( \hat{r}^4 \sin^2\theta \partial_{\hat{r}}\hat{A}^{\phi}\right)+\frac{1}{\sin\theta}\partial_{\theta} \left(\sin^3\theta\,\partial_{\theta} \hat{A}^{\phi}\right)\approx \hat{\chi} \hat{\chi}^*.
\end{equation}

\section{Solution}
\label{Sec:Sol}
\subsection{A Simple WKB Analysis\label{eq:secWKB}}
For large values of $m_{\phi}$, the profile for $\hat{\chi}$ localises at large $\hat{r}$. This is because the centrifugal force causes $\hat{\chi}$ to move outwards. At this point, $\hat{\chi}$ is essentially vanishing near the origin $\hat{r}\approx0$. Recall the approximate radion equation of motion
\begin{equation}
	\nabla^2 \delta \hat{a} = 4\pi \hat{\chi} \hat{\chi}^*.
 \label{eq:simphata}
\end{equation}
Integrating both sides of the above equation, and using the normalisation for $\hat{\chi}$ implies that
\begin{equation}
\delta \hat{a}\approx -\frac{1}{\hat{r}}\,
\end{equation}
at large values of $\hat{r}$. This is essentially the potential that $\hat{\chi}$ feels at large $\hat{r}$. The equation for $\hat{\chi}$, for large values of $m_{\phi}$ can then be written as
\begin{equation}
\frac{1}{\hat{r}^2}\partial_{\hat{r}}\left(\hat{r}^2\partial_{\hat{r}} \hat{\chi}\right)+\frac{1}{\hat{r}^2\sin\theta}\partial_{\theta}(\sin \theta \partial_{\theta}\hat{\chi})-\frac{m_{\phi}^2}{\hat{r}^2\sin^2\theta}\hat{\chi}-\left(\zeta-\frac{1}{\hat{r}}\right) \hat{\chi}\approx0\,.
\end{equation}
The above equation is essentially a Schrödinger equation for a $1/\hat{r}$ Coulomb potential, which can be readily solved using separation of variables by setting
\begin{equation}
\hat{\chi}=R(\hat{r})P^{m_{\phi}}_{\ell}(\cos \theta)~,
\end{equation}
with $P^{m_{\phi}}_{\ell}(z)$ an associated Legendre polynomial of degree $\ell\geq |m_{\phi}|$ and order $m_{\phi}$. From the above we find
\begin{equation}
\frac{1}{\hat{r}^2}\partial_{\hat{r}}\left(\hat{r}^2\partial_{\hat{r}} R\right)-\frac{\ell(\ell+1)}{\hat{r}^2}R(\hat{r})-\left(\zeta-\frac{1}{\hat{r}}\right) R(\hat{r})\approx0\,.
\end{equation}
The general solution to the above ordinary differential equation reads
\begin{multline}
R(\hat{r})=e^{-\hat{r}\sqrt{\zeta}}\hat{r}^{\ell}\Big[C_1 U\left(\ell-\frac{1}{2 \sqrt{\zeta }}+1;2 \left(\ell+1\right);2 \sqrt{\zeta } \hat{r}\right)
\\
+C_2\,L\left(-\ell+\frac{1}{2 \sqrt{\zeta }}-1;2 \ell+1;2 \sqrt{\zeta } \hat{r}\right)\Big]\,
\label{eq:Rhat}
\end{multline}
where $U(a;b;z)$ is the Tricomi confluent hypergeometric function, and $L(a;b;z)$ is the generalized Laguerre function. Demanding that $\hat{\chi}$ vanishes as $\hat{r}\to+\infty$ selects $C_2=0$.

At the origin we impose that $\hat{\chi}$ is finite (indeed, we expect $\hat{\chi}$ to vanish near $\hat{r}\simeq0$). Expanding $R(\hat{r})$ at small $\hat{r}$, and demanding that the term proportional to $\hat{r}^{-1-2\ell}$ is altogether absent, quantises $\zeta$ to be
\begin{equation}
\zeta \approx \frac{1}{4(1+p+\ell)^2}\quad\text{with}\quad p\in\mathbb{N}_0\quad \text{and}\quad \ell\geq |m_{\phi}|\,,
\end{equation}
where the approximate sign reminds us that the above is only valid for large enough $|m_{\phi}|$. The ground state corresponds to $p=0$ and $\ell=|m_{\phi}|$, and we thus expect the ground state value of $\zeta$ to behave as
\begin{equation}
\zeta \approx \frac{1}{4(1+|m_{\phi}|)^2}\,.
\label{eq:zetaapprox}
\end{equation}
at large values of $m_{\phi}$. Our numerical methods will confirm the above expectations. One can do a little better and also work out what happens in the near zone, i.e. for $\hat{r}\ll 1$. Since we expect the scalar field to be localised at large $\hat{r}$, we can neglect the right-hand-side in the equation for $\delta \hat{a}$. This means that in this region $\delta \hat{a}\approx -a_0$, with $a_0$ being a constant to be determined in what follows. Using this we can solve for $\hat{\chi}$ in terms of spherical Bessel functions of the first and second kind, namely
\begin{equation}
\label{SmallR}
\hat{\chi}(\hat{r},\theta)=P_{\ell}^{m_{\phi}}(\theta) \left[C_3 \,j_{\ell}\left(\hat{r} \sqrt{a_0-\zeta }\right)+C_4\,y_{\ell}\left(\hat{r} \sqrt{a_0-\zeta }\right)\right]\,
\end{equation}
where $C_3$ and $C_4$ are constants of integration, and $j_{\ell}$ and $y_{\ell}$ denote the spherical Bessel functions of the first and second kind, respectively. Regularity at the origin $\hat{r}=0$ fixes $C_4=0$. If our approximation is to be valid, the large $\hat{r}$ behavior of $\hat{\chi}(\hat{r},\theta)$ in Eq.~(\ref{eq:Rhat}) with $C_2=0$ should match the small $\hat{r}$ behavior in (\ref{SmallR}) with $C_4=0$. This will only be the case if $a_0=\zeta$. We thus conclude that, in the large $|m_{\phi}|$ limit,
\begin{equation}
\delta \hat{a}(0,\theta)\approx -\zeta\,,
\end{equation}
and in particular, for the ground state $\ell=m_{\phi}$ and $p=0$, we find
\begin{equation}
\delta \hat{a}(0,\theta)\approx  -\frac{1}{4(1+|m_{\phi}|)^2}\,.
\label{eq:deltaaapprox}
\end{equation}
We shall see later, using accurate numerical methods, that (\ref{eq:deltaaapprox}) yields an excellent approximation for the behaviour of $\delta\hat{a}$ near the origin as $|m_{\phi}|$ grows large.

In principle, one could go even further and take the leading behaviour of $\delta \hat{a}$ near infinity, and substitute again in the equation for $\hat{\chi}$.

\subsection{The Self Gravitating Rotating String }
We use numerical techniques to solve the coupled system of partial differential equations composed of Eq.~(\ref{ChiEqfinal?}) and Eq.~(\ref{RadionEq}). Recall that for $\hat{\chi}$ we give a definite phase of the form
\begin{equation}
\hat{\chi}(\hat{r},\theta,\phi)=e^{i m_{\phi} \phi}\tilde{\chi}(\hat{r},\theta)\,.
\end{equation}
while $\delta \hat{a}$ is a function of $(\hat{r},\theta)$ only. We solve the above system of equations using spectral Galerkin methods \footnote{We have also solved the same equations using Chebyshev collocation methods. However, we found that with the latter it was very hard to find solutions for $m_{\phi}\geq5$. This is to contrast with the spectral Galerkin method detailed in the main text for which we reached $m_{\phi}=8$.}. The method is based on a \emph{natural} expansion of $\delta \hat{a}$ and $\hat{\chi}$ in terms of Legendre and associated Legendre polynomials, respectively. In particular, we set
\begin{subequations}
\begin{equation}
\delta \hat{a}(\hat{r},\theta)=\sum_{\ell=0}^{+\infty} P_{\ell}(\cos \theta) \hat{a}_{\ell}(\hat{r})
\end{equation}
and
\begin{equation}
\hat{\chi}(\hat{r},\theta)=\sum_{\ell=m_{\phi}}^{+\infty} P^{m_{\phi}}_{\ell}(\cos \theta) \hat{\chi}_{\ell}(\hat{r})
\end{equation}
\end{subequations}%
where $P_{\ell}$ and $P^{m_{\phi}}_{\ell}$ are Legendre polynomials of degree $\ell$ and associated Legendre polynomials of degree $\ell$ and order $m_{\phi}$. The expansion coefficient functions $\hat{\chi}_{\ell}(\hat{r})$ and $\hat{a}_{\ell} (\hat{r})$ are taken to be real. The above expansions automatically ensure smoothness of $\delta \hat{a}$ and $\hat{\chi}$ in the polar directions. We can input these expansions in Eq.~(\ref{ChiEqfinal?}) and Eq.~(\ref{RadionEq}), and project each of these equations into a given $\ell$ mode. The final result depends on an infinite double sum of an integral involving the product of three Legendre and associated Legendre polynomials that can be solved analytically in terms of $3-j$ symbols. The resulting infinite set of equations read
\begin{subequations}
\begin{multline}
\frac{1}{\hat{r}^2}\partial_{\hat{r}}\left(\hat{r}^2\partial_{\hat{r}} \hat{a}_{\ell}\right)-\frac{\ell(\ell+1)}{\hat{r}^2}\hat{a}_{\ell}
\\
-4\pi \,(-1)^{m_{\phi}}\,\left(2\ell+1\right)\sum_{\hat{\ell}=m_{\phi}}^{+\infty}\sum_{\tilde{\ell}=m_{\phi}}^{+\infty}\sqrt{\frac{(\hat{\ell}+m_{\phi})!(\tilde{\ell}+m_{\phi})!}{(\hat{\ell}-m_{\phi})!(\tilde{\ell}-m_{\phi})!}}\left(\begin{array}{ccc}
\ell & \hat{\ell} & \tilde{\ell} \\ 0 & 0 &0\end{array}\right)\left(\begin{array}{ccc}
\ell & \hat{\ell} & \tilde{\ell} \\ 0 & m_{\phi} &-m_{\phi}\end{array}\right)\hat{\chi}_{\hat{\ell}}\hat{\chi}_{\tilde{\ell}}=0\,,
\end{multline}
and
\begin{multline}
\frac{1}{\hat{r}^2}\partial_{\hat{r}}\left(\hat{r}^2\partial_{\hat{r}} \hat{\chi}_{\hat{\ell}}\right)-\frac{\hat{\ell}(\hat{\ell}+1)}{\hat{r}^2}\hat{\chi}_{\hat{\ell}}-\zeta\,\hat{\chi}_{\hat{\ell}}
\\
-(-1)^{m_{\phi}}\,\left(2\hat{\ell}+1\right)\sqrt{\frac{(\hat{\ell}-m_{\phi})!}{(\hat{\ell}+m_{\phi})!}}\sum_{\ell=0}^{+\infty}\sum_{\tilde{\ell}=m_{\phi}}^{+\infty}\sqrt{\frac{(\tilde{\ell}+m_{\phi})!}{(\tilde{\ell}-m_{\phi})!}}\left(\begin{array}{ccc}
\hat{\ell} & \ell & \tilde{\ell} \\ 0 & 0 &0\end{array}\right)\left(\begin{array}{ccc}
\hat{\ell} & \ell & \tilde{\ell} \\ m_{\phi} & 0 &-m_{\phi}\end{array}\right)\hat{a}_{\ell}\hat{\chi}_{\tilde{\ell}}=0\,,
\end{multline}
where
\begin{equation}
\left(\begin{array}{ccc}
\ell_1 & \ell_2 & \ell_3 \\ m_1 & m_2 &m_3\end{array}\right)
\end{equation}
\end{subequations}%
is the standard Wigner $3-j$ symbol and $\ell=0,1,\ldots$, while $\hat{\ell},\tilde{\ell}=m_{\phi},m_{\phi}+1,\ldots$. Numerically, we will truncate the above infinite sum to a finite number of terms. In particular, we will take $\ell=0,1,\ldots,2(m_{\phi}+N_{\theta})$ and $\hat{\ell},\tilde{\ell}=m_{\phi},m_{\phi}+1,\ldots,m_{\phi}+N_{\theta}$. A \emph{posteriori}, we can determine the convergence of the numerical method by dialing for several values of $N_{\theta}$. The above method is similar in spirit to Galerkin spectral methods and we expect exponential convergence in $N_{\theta}$ to the continuum limit. We are left with a \emph{finite} system of ordinary differential equations in $\{\hat{a}_{\ell},\hat{\chi}_{\hat{\ell}}\}$. To solve these, we introduce a compact coordinate $y$ via the relation
\begin{equation}
\hat{r}=\frac{y}{1-y}\Rightarrow y =\frac{\hat{r}}{1+\hat{r}}\,,
\label{eq:maprh}
\end{equation}
with the origin ($\hat{r}=0$) being located at $y=0$, and asymptotic infinity ($\hat{r}\to+\infty$) at $y=1$. Finally, we also perform a change of variable in which we take
\begin{equation}
\hat{a}_{\ell}=y^{\ell}q^{(1)}_{\ell}\quad\text{and}\quad \hat{\chi}_{\hat{\ell}}=(1-y)^2y^{\ell}q^{(2)}_{\hat{\ell}}\,,
\end{equation}
and solve for $\{q^{(1)}_{\ell},q^{(2)}_{\hat{\ell}}\}$.

At the origin, located at $y=0$, we impose regularity, which in turn implies
\begin{equation}
{q^{(1)}_{\ell}}^\prime(0)-\ell q^{(1)}_{\ell}(0)=0\quad\text{and}\quad{q^{(2)}_{\hat{\ell}}}^\prime(0)-(\hat{\ell}+2) q^{(2)}_{\ell}(0)=0\,.
\end{equation}
On the other hand, at asymptotic infinity, we demand
\begin{equation}
{q^{(1)}_{\ell}}(1)={q^{(2)}_{\hat{\ell}}}(1)=0\,.
\end{equation}
Recall that we also want to impose a normalisation condition on $\hat{\chi}$, so that
\begin{subequations}
\begin{equation}
    \int_0 ^{\infty} {\rm d}\hat{r} \int_0 ^{\pi} {\rm d}\theta \int_0 ^{2\pi} {\rm d}\phi ~\hat{r}^2\,\sin\theta\,|\hat{\chi}(\hat{r},\theta)|^2=1
\end{equation}
which in turn translates into a normalisation condition involving all the $q^{(2)}_{\hat{\ell}}$
\begin{equation}
    \sum_{\ell=m_{\phi}}^{+\infty}\frac{(\ell+m_{\phi})!}{(\ell-m_{\phi})!}\frac{1}{2\ell+1}\int_0 ^{1} {\rm d}y~y^{2\ell+2}{q^{(2)}_{\ell}}^2=\frac{1}{4\pi}\,.
\end{equation}
\end{subequations}
To solve for $q_{\ell}^{(1)}$ and $q_{\hat{\ell}}^{(2)}$, we discretise the system on a Gauss-Lobatto collocation grid with $N_y$ points. The resulting equations are then solved using relaxation methods along with a standard Newton-Raphson routine. For further details on these methods, we refer the readers to \cite{Dias:2015nua}. Since $q_{\hat{\ell}}^{(2)}$ is highly localized near $y=1$, we employ a rather dense grid with $N_y$ typically in the range of $2\times 10^2$ to $10^{3}$. In appendix \ref{app:conv} we provide a short convergence survey.

\begin{figure}[t]
    \centering\includegraphics[height=6cm]{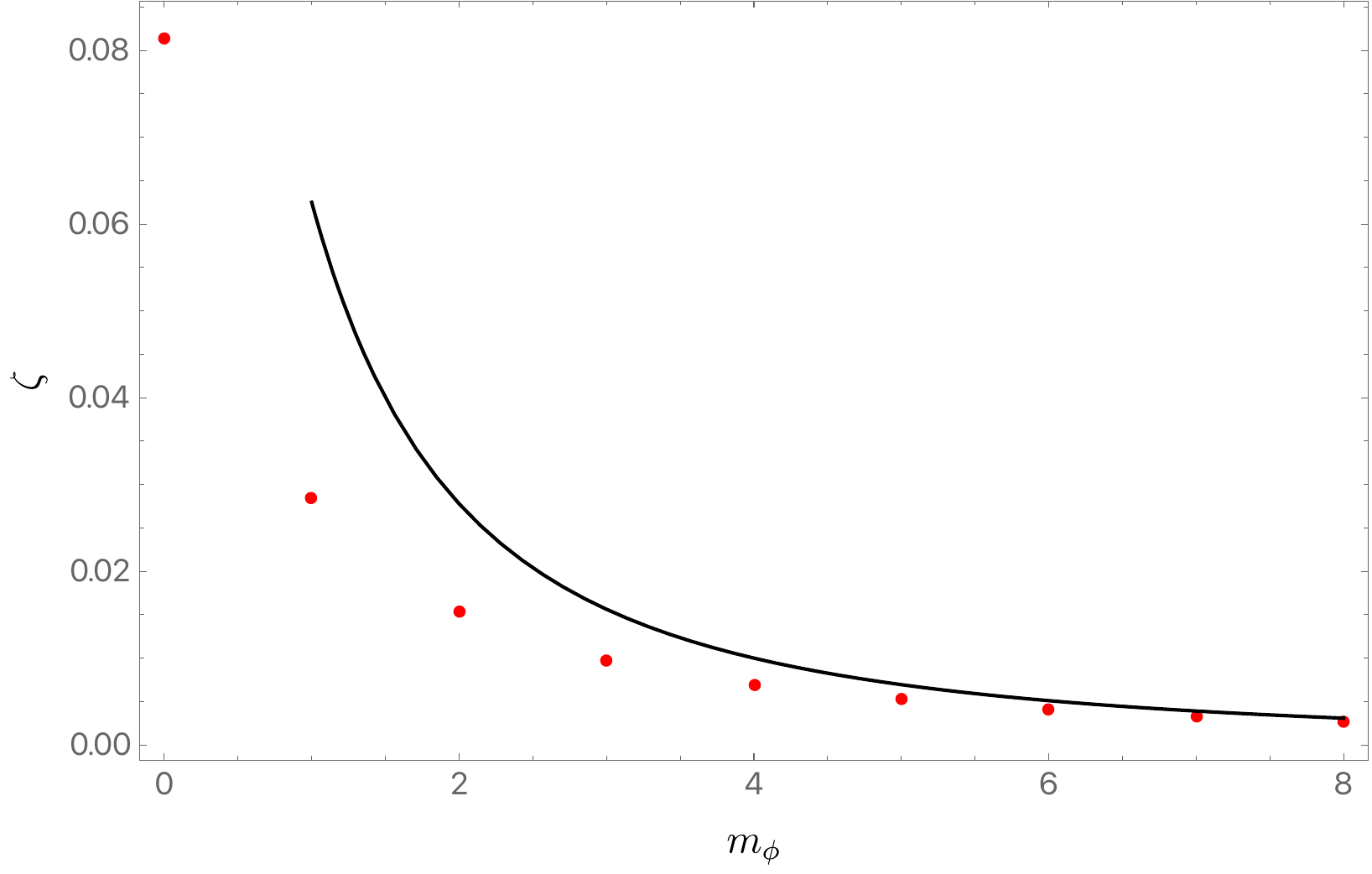}
    \caption{$\zeta$ as a function of $m_{\phi}$: the red disks represent our exact numerical data, while the solid black line is given by (\ref{eq:zetaapprox}).}
    \label{fig:zeta}
\end{figure}

Our numerical results for $\zeta$ are plotted in Fig.~\ref{fig:zeta}, where we show $\zeta$ for $m_{\phi}=1,\ldots,8$. The red disks represent our exact numerical data, while the solid black line is given by (\ref{eq:zetaapprox}). For large enough $m_{\phi}$, the agreement between our numerical scheme and the analytic approximation valid at large $m_{\phi}$ is reassuring.

\begin{figure}[t]
    \centering\includegraphics[height=6cm]{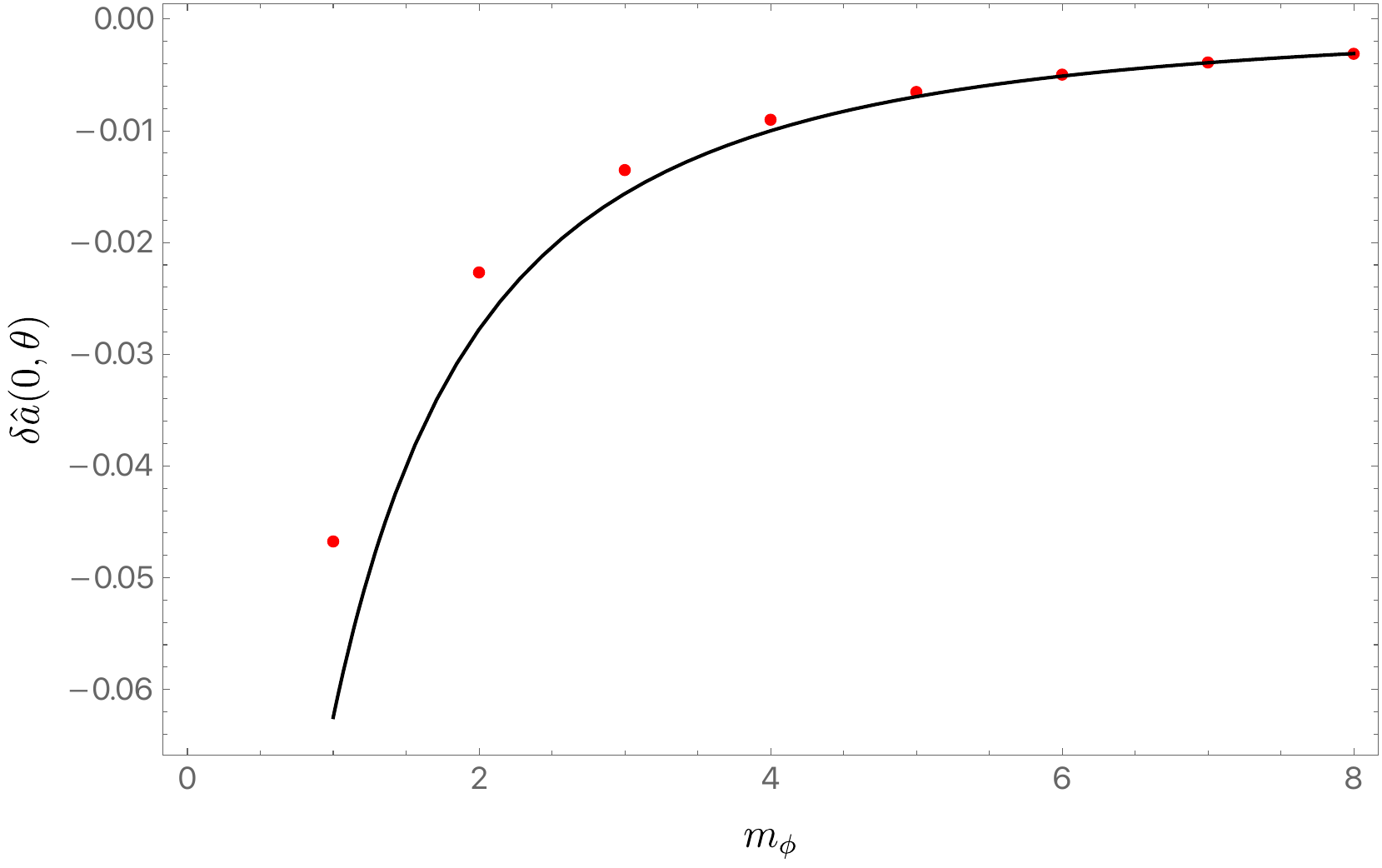}
    \caption{$\delta\hat{a}(0,\theta)$ as a function of $m_{\phi}$: the red disks represent our exact numerical data, while the solid black line is given by (\ref{eq:zetaapprox}).}
    \label{fig:deltaa}
\end{figure}

In Fig.~\ref{fig:deltaa}, we present $\delta\hat{a}(0,\theta)$ plotted against $m_{\phi}$. Similar to Fig.~\ref{fig:zeta}, the exact numerical data is depicted as red disks, while the analytic expansion (\ref{eq:deltaaapprox}) is represented by a solid black line. Once more, we observe an excellent agreement at sufficiently large $m_{\phi}$ values

\begin{figure}[t]
    \centering\includegraphics[height=6cm]{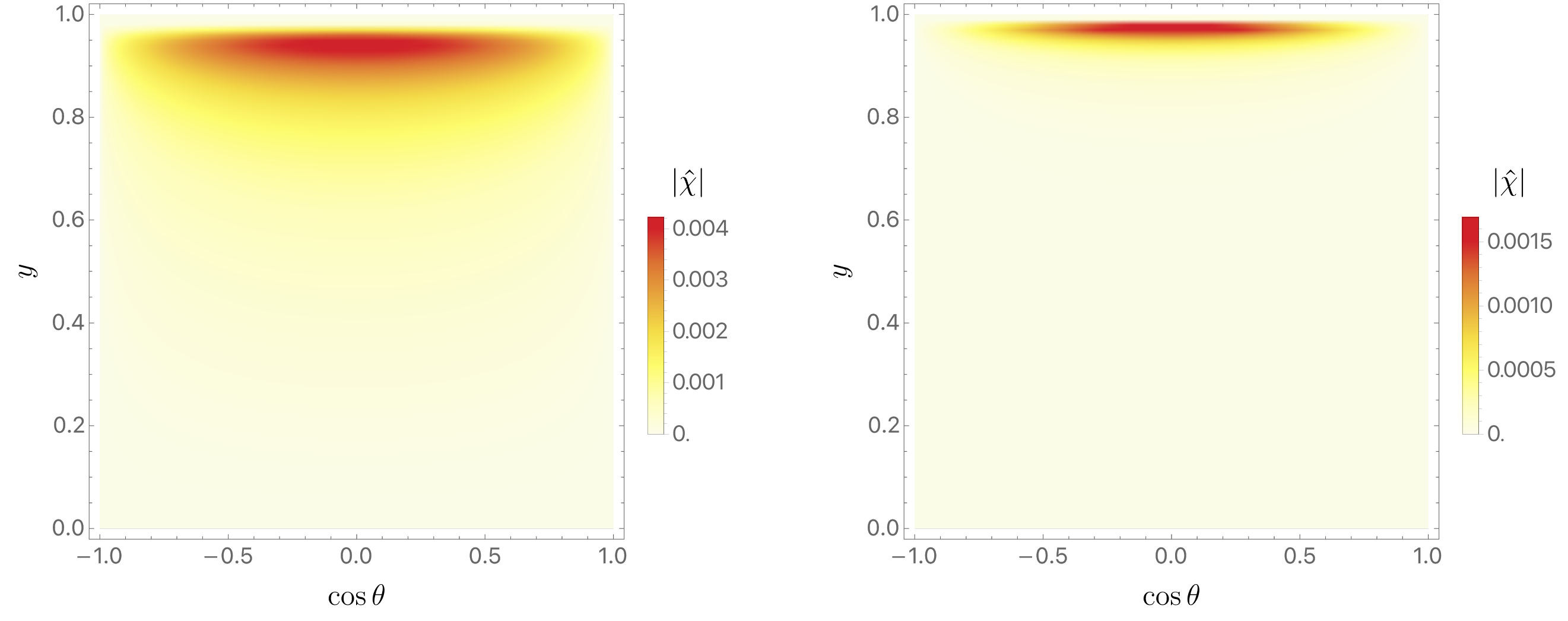}
    \caption{$|\hat{\chi}|$ as a function of $y$ and $\cos \theta$ for $m_{\phi}=1$ (left panel) and $m_{\phi}=2$ (right panel).}
    \label{fig:chi}
\end{figure}

In Fig.~\ref{fig:chi}, we illustrate $|\hat{\chi}|$ as a function of $y$ and $\cos \theta$ for two scenarios: $m_{\phi}=1$ (left panel) and $m_{\phi}=2$ (right panel). Notably, with increasing $m_{\phi}$, $|\hat{\chi}|$ tends to localise closer to $\theta = \pi/2$ and $y=1$, aligning with our expectations from the simple WKB approximation. It is worth mentioning that the $|\hat{\chi}|$ profile peaks at large values of $\hat{r}$, allowing us to approximate $\hat{a}$ in this region as $-1/\hat{r}$. Finally, we also point out that the accumulation of the signal near $y=1$ as $m_{\phi}$ increases appears very drastic. However, this is partially a consequence of the fact that the mapping between $\hat{r}$ and $y$, i.e., Eq.~(\ref{eq:maprh}), is such that almost all of spatial infinity gets squeezed near $y=1$.
\subsection{Thermodynamic Quantities}
In this subsection, we compute the size, mass, angular momentum, entropy, and winding charge of the self-gravitating spinning string condensate. 

The size of the solution scales like the inverse of the mass of the condensate:
\begin{equation}
	D \sim \frac{1}{m} \propto \frac{\left(\alpha'\right) ^{\frac{3}{4}}}{\sqrt{\beta - \beta_H}}.
\end{equation}

To compute the energy, we simply evaluate the ADM energy\footnote{Alternatively, we could have used the Komar energy at spatial infinity, since the solution is stationary.}. In our approximation, the energy reads
\begin{equation}
\label{Mass}
	M\approx \frac{ \alpha' m }{2 G_N\kappa_{het}\sqrt{\zeta}}.
\end{equation}

The angular momentum of the self-gravitating string $J$ can also be computed via the ADM formalism by reading the asymptotic falloff near spatial infinity of the graviphoton field: 
\begin{equation} 
	G_{\tau \phi}=-r^2 \sin^2\theta \omega(r,\theta) \approx -\frac{\nu \sin^2\theta}{r^{D-3}} ~,~ 
\end{equation}
\begin{equation}
\label{Jsol}
    J=\frac{\omega_{D-2}~ \nu}{ 8\pi G_N}.
\end{equation}
In particular, for $D=4$, $\omega(r,\theta)$ admits the asymptotic form
\begin{equation}
	\omega(r,\theta)\to \frac{\nu}{r^{3} }.
\end{equation}
Eq.~(\ref{OmegaScaling}) for the scaling of $\omega$ and $r\propto \frac{1}{m}$ imply that
\begin{equation}
	\nu =\frac{(4\pi)^2 m_{\phi} (\alpha')^2 m }{\beta\kappa_{het}^2 \sqrt{\zeta}} \hat{\nu}~~,~~     \hat{\nu} = \lim _{\hat{r}\to \infty} \hat{r}^{3} \hat{\omega}(\hat{r},\theta). 
\end{equation}
Later in this subsection, we determine $\hat{\nu}$.
From Eq.~(\ref{Jsol}) it follows that
\begin{equation}
\label{AngularMomentum1}
	J = \frac{\nu }{ 2G_N}= \frac{8\pi^2 {\alpha^\prime}^2  m m_{\phi}}{G_N \beta\kappa_{het}^2 \sqrt{\zeta}} \hat{\nu}.
\end{equation}
Another approach for computing the angular momentum is to use a thermodynamics relation. If one works in an ensemble of fixed angular velocity in which states are weighted by $\propto \frac{1}{Z}e^{-\beta \Omega \hat{J}}$, then the expectation value of the angular momentum is given by 
\begin{equation}
	J =- \frac{1}{\beta} \frac{\partial \log Z}{\partial \Omega} =\frac{1}{\beta}\frac{\partial I_D }{\partial \Omega}.
\end{equation}
We have defined the angular velocity of the object as the value of $\omega(r,\theta)$ at the origin $\Omega \equiv \omega(0,\theta)$. 
Before evaluating the on-shell action, one can note that the action contains a term
that comes from multiplying the graviphoton field that appears in the covariant derivative of $\chi$ and the current associated with $U(1)$ gauge redundancy:
\begin{equation}
	\label{AJcoupling}
	I \supset \beta \int {\rm d}^d x \sqrt{G_D} e^{-2\Phi_D} \frac{\pi}{\beta}A_{\phi} j^{\phi} \approx 2\pi \int {\rm d}^d x \sqrt{G_D}e^{-2\Phi_D} \left(-\frac{\omega}{a}\right)m_{\phi}\chi \chi^*.
\end{equation}
We have used $A_{\phi}\approx \frac{1}{a} r^2 \sin^2\theta \omega (r,\theta)$ and $j_{\phi}\approx 2m_{\phi} \chi \chi^*$. 
Setting $d=3$ and taking a partial derivative with respect to $\Omega$ while holding $\beta$ and the chemical potential for winding charge fixed give rise to 
\begin{equation}
	J \approx 2 \frac{\pi}{\beta} m_{\phi} \int {\rm d}r\,{\rm d}\theta\,{\rm d}\phi\, r^2 \sin\theta  \chi \chi ^*=8\pi^2 m_{\phi} \frac{(\alpha' m^2)^2}{\kappa_{het}^2 \zeta^2 \times 8\pi G_N} \frac{\zeta^{\frac{3}{2}}}{m^3}  \int {\rm d}\hat{r}\,{\rm d}\theta\,{\rm d}\phi\, \hat{r}^2 \sin\theta  \hat{\chi} \hat{\chi} ^*.
\end{equation}
Consistency with Eq.~(\ref{AngularMomentum1}) implies that
\begin{equation}
	\hat{\nu} =\frac{1}{8\pi}\int {\rm d}\hat{r}\,{\rm d}\theta\,{\rm d}\phi\,  \hat{r}^2 \sin\theta  \hat{\chi} \hat{\chi} ^*.
\end{equation}
As a check, one can directly integrate the Poisson equation (\ref{BoxA}) to find $\hat{\nu}$:
\begin{equation}
	\int {\rm d}\hat{r}\,{\rm d}\theta\,{\rm d}\phi ~\partial_{\hat{r}} \left[ \hat{r}^2 \sin \theta\,\hat{r}^2 \sin^2\theta  (\partial^{\hat{r}}\hat{A} ^{\phi})\right]=	\int {\rm d}\hat{r}\,{\rm d}\theta\,{\rm d}\phi\,\hat{r}^2 \sin \theta \hat{\chi} \hat{\chi} ^*.
\end{equation}
Since $\hat{A}_{\phi} \approx -\hat{\omega} \hat{r}^2 \sin^2\theta$, $\hat{A}^{\phi} \approx -\hat{\omega}(\hat{r},\theta)$. Writing $\hat{\omega} = \frac{\hat{\nu}}{\hat{r}^3}$ asymptotically results in
\begin{equation}
	\hat{\nu} = \frac{1}{8\pi}\int {\rm d}\hat{r}\,{\rm d}\theta\,{\rm d}\phi \,\hat{r}^2 \sin \theta \hat{\chi} \hat{\chi} ^*.
\end{equation}
The normalization condition on $\chi$ simplifies the resulting angular momentum to be
\begin{equation}
\label{AngularMomentum2}
	J \approx  \frac{\pi {\alpha^\prime}^2 \,m\,m_{\phi}}{G_N \beta\,\kappa_{het}^2 \sqrt{\zeta}} =\frac{2\pi m_{\phi}}{\kappa_{het} \beta} \alpha^\prime M\,,
\end{equation}
from which we conclude that, 
\begin{equation}
g_s^2J\approx\frac{1}{16\ 2^{5/4}
   \left(1+\sqrt{2}\right) \pi}\frac{m_{\phi}}{\sqrt{\zeta}}\sqrt{\frac{\beta -\beta _H}{\beta _H}}\,,
\end{equation}
where we used that $2\pi \alpha^\prime=\ell_s^2$ and $G_N=g_s^2\,\ell_s^2$ in four spacetime dimensions\footnote{We have used the relation $G_N^{(10D)}=8 \pi^6 g_s^2 {\alpha^\prime}^4$ \cite{Polchinski2}, and for the sake of clarity in presentation, we compactify to four spacetime dimensions by assuming that the remaining six-torus possesses a total volume of $\pi^2 \ell_s^6/2$.}. 

\begin{figure}[t]
    \centering\includegraphics[height=6cm]{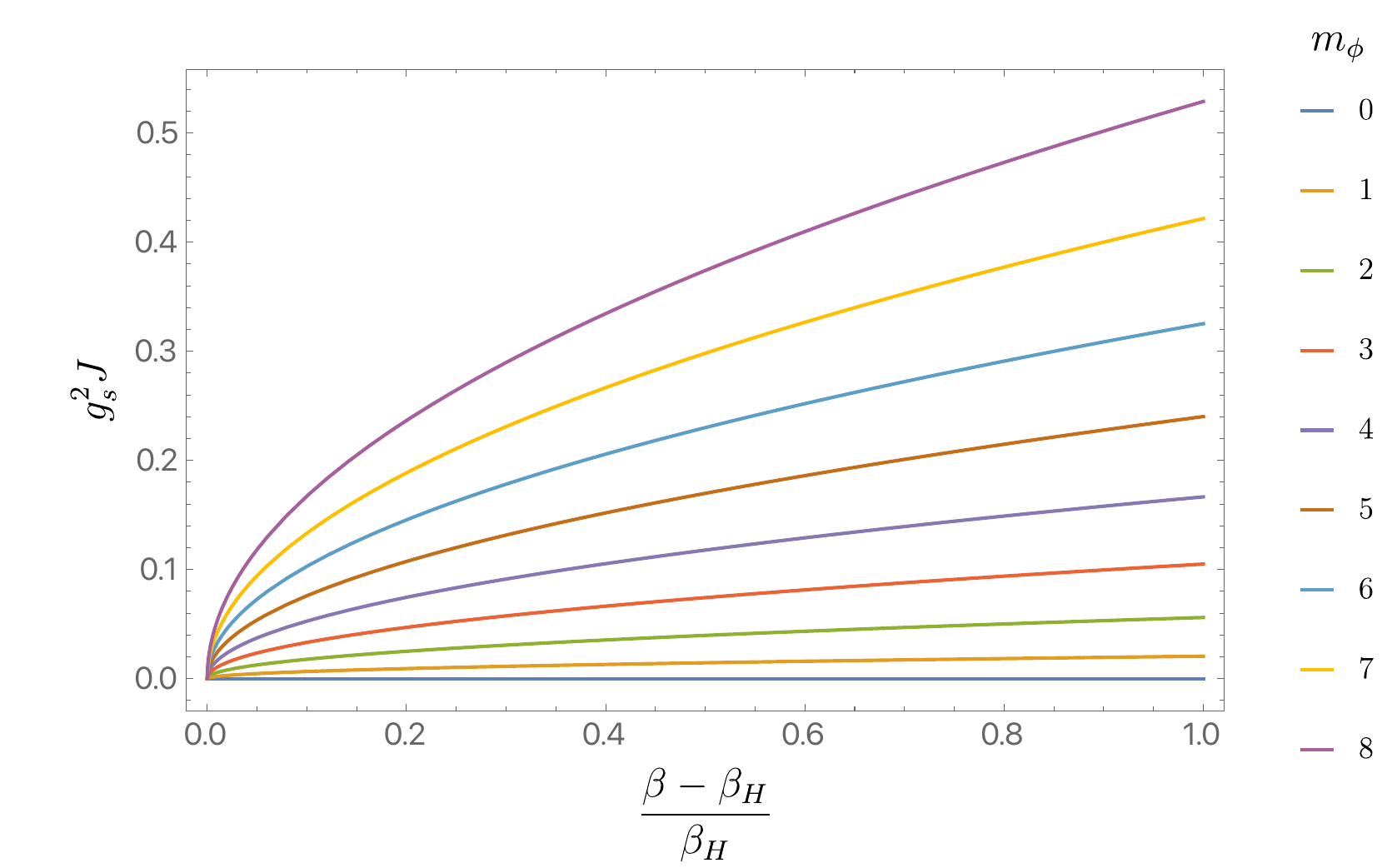}
    \caption{The rescaled angular momentum $g_s^2 J$ is depicted as a function of $(\beta-\beta_H)/\beta_H$ for several fixed values of $m_{\phi}$ labeled on the right.}
    \label{fig:angular}
\end{figure}

In Fig.~\ref{fig:angular} we plot $g_s^2 J$ as a function of $(\beta-\beta_H)/\beta_H$ for several values of $m_{\phi}$.
As expected, the angular momentum grows faster as a function of $(\beta-\beta_H)/\beta_H$ with increasing values of $m_{\phi}$. As the temperature decreases, the angular momentum increases. Intuitively, this is explained by the decreasing trend of the moment of inertia of the winding-momentum condensate, $I\sim M D^2 \sim \frac{1}{\sqrt{\beta-\beta_H}}$, implying that at lower temperatures, the condensate resistance to spin decreases. 

For a given Euclidean action evaluated at fixed $\beta$, chemical potential $\mu$, and angular velocity $\Omega$, the entropy is given by
\begin{equation}
	S = \left(\beta \left.\frac{\partial}{\partial \beta}\right|_{\Omega,\mu} -1\right)I_D = \beta^2 \left.\frac{\partial I_d}{\partial \beta}\right|_{\Omega,\mu}\,,
\end{equation}
where in the last equality we have used $I_D = \beta I_d$. To compute the above, we recall that $I_d$ is a functional of the $d$-dimensional metric, of the graviphoton $A_{\mu}$, $\sigma$, $\chi$, $B_{\mu \tau}$, $B_{\mu \nu}$ and $\Phi_d$. Let us denote these fields collectively as $\Phi^I$. In general, the $\Phi^I$ will depend on $\beta$. Assuming that a standard Gibbons-Hawking-York boundary term is added to make the variational problem well defined \cite{York:1972sj,Gibbons:1976ue}, $\left.\frac{\partial I_d}{\partial \beta}\right|_{\Omega,\mu}$ picks up two types of contributions:
\begin{multline}
\left.\frac{\partial I_d[\Phi^I(\beta,\Omega,\mu),\beta,\Omega,\mu]}{\partial \beta}\right|_{\Omega,\mu}=\frac{\delta I_d[\Phi^I(\beta,\Omega,\mu),\beta,\Omega,\mu]}{\delta \Phi^J(\beta,\Omega,\mu))}\left.\frac{\partial \Phi^J(\beta,\Omega,\mu)}{\partial \beta}\right|_{\Omega,\mu}
\\
+\left.\frac{\partial I_d[\Phi^I(\beta,\Omega,\mu),\beta,\Omega,\mu]}{\partial \beta}\right|_{\Phi^I(\beta,\Omega,\mu),\Omega,\mu}\,.
\end{multline}
The first term vanishes because it is proportional to the equations of motion for $\Phi^I$. The last term is computed by taking derivatives of the action with respect to \emph{explicit} dependencies on $\beta$. There are two such contributions, both of which stemming from $I_1/\beta$, where $I_1$ is defined in Eq.~(\ref{eq:I1N}). The first contribution is a consequence of the fact that $m$ explicitly depends on $\beta$ - see Eq.~(\ref{eq:m2}). The second contribution comes from the covariant derivatives acting on $\chi$ which also depend explicitly on $\beta$. Once the dust settles, we find
\begin{multline}
S\approx\frac{\beta_H \alpha^\prime m }{2 G_N\kappa_{het}\sqrt{\zeta}}
\\
-\frac{\beta_H  m^5 m_{\phi }^2 \pi^2 {\alpha^\prime}^3}{2  G_N \kappa_{het}\zeta^{5/2}}\int_0^{\infty}\mathrm{d}\hat{r}\int_{0}^{\pi} \mathrm{d}\theta\,\hat{r}^2 \sin \theta\, |\hat{\chi}(\hat{r},\theta)|^2 \hat{\omega}\,(\hat{r},\theta)\Big[1-\frac{3 \pi \alpha^\prime m^2 }{2 \zeta } \hat{r}^2 \sin ^2\theta\,\hat{\omega}(\hat{r},\theta)\Big]\,,
\label{eq:totalentropy}
\end{multline}
where we used that $\hat{A}_{\phi}=\hat{B}_{\tau\phi}$, $\hat{A}_{\phi}=-\hat{r}^2\,\sin^2\theta \,\hat{\omega}$ and imposed the usual normalisation for $\hat{\chi}$ in $d=3$, namely
\begin{equation}
    1 = \int_0 ^{\infty} {\rm d}\hat{r} \int_0 ^{\pi} {\rm d}\theta \int_0 ^{2\pi} {\rm d}\phi ~\hat{r}^2 \sin\theta\,|\hat{\chi}(\hat{r},\theta)|^2.
\end{equation}
Our expression for the entropy (\ref{eq:totalentropy}) may be misleading. Specifically, it could erroneously suggest that there are no corrections between order $m$ and $m^5$. However, this is certainly not the case. For instance, in \cite{CMW}, corrections proportional to $m^3$ were found. Such corrections are expected to be more significant compared to those arising from the second line in Eq.~(\ref{eq:totalentropy}), if considering moderate values of $m_{\phi}$. Instead, we should interpret the corrections appearing on the second line as being the leading corrections to the entropy generated \emph{via rotation}. In fact, we note that the first term on the second line in Eq.~(\ref{eq:totalentropy}) takes roughly the form of $J \Omega$, which is the work term expected for a rotating system.

We will focus on a regime of parameters where the second line in Eq.~(\ref{eq:totalentropy}) is subdominant with respect to the first. Since $\zeta\sim m_{\phi}^{-2}$ at large $m_{\phi}$, this will only occur if $m_{\phi}\ll (\ell_s m)^{-2/3}$. This is the parameter regime we will henceforth consider, and for which the entropy is simply
\begin{equation}
S\approx\frac{\beta_H \alpha^\prime m }{2 G_N\kappa_{het}\sqrt{\zeta}}\,.
\end{equation}
It is a simple exercise to check that, to leading order in $m$, the relation $\beta = \left(\frac{\partial S}{\partial M}\right)_{J}$ together with $\beta \approx \beta_H$, imply that the energy of the solution is given as in Eq.~(\ref{Mass}).

The NS-NS winding charge can be found utilising Eq.~(\ref{FvsH}), which expresses the fact that we are discussing a bound state that carries $\frac{\pi}{\beta}$ units of momentum for any $\frac{\beta}{2\pi \alpha'}$ units of winding:
\begin{equation}
Q = \frac{\beta^2}{2\pi^2 \alpha'} J =\frac{\alpha' \beta  m m_{\phi}}{2\pi G_N \kappa_{het}^2 \sqrt{\zeta}}\,.
\end{equation}

At first, it might appear puzzling that the self-gravitating strings carry NS-NS charge. However, the topology of our solution is such that the $B$ field can wrap a nontrivial two-cycle, namely the two-torus spanned by $(\tau,\phi)$. Note, however, that there are no homologically non-trivial three-manifolds (\emph{e.g.}, $\mathbb{T}^3$), and as such, the charge need not be quantized \cite{Bowick:1988xh}.

\subsection{Low Temperature Phase}

We now imagine that the temperature is lowered so that it is not approximately the Hagedorn value and that the charges $J,Q$ are fixed. The winding mode becomes heavy so that it is outside the effective field theory of the heterotic string. An expectation for the nature of the low-temperature phase is a version of the Kerr black hole rotating in Euclidean time which carries NS-NS string charge. Without the additional charge, a rotating black hole metric with respect to Euclidean time has an interpretation as a pair of KK monopoles \cite{GrossPerry83}, where the fluxes produced by the magnetic monopoles solve the relevant supergravity equations. This follows from interpreting the graviphoton as a vector potential for the magnetic field, which asymptotically $A_{\phi} \propto \frac{\sin^2 \theta}{r^2}$ can be explained by a dipole of magnetic monopole and anti-monopole.
In our case, the additional NS-NS charge implies  ``dyonic monopoles'' that couple to the $A$- and $B$-fields. 

A static black hole solution of mass $M$ with axion charge $Q$ was first written by \cite{Bowick:1988xh}
\begin{equation}
\label{BGHHS}
    {\rm d}s^2 = -\left(1-\frac{2M}{r}\right){\rm d}t^2 + \frac{{\rm d}r^2}{1-\frac{2M}{r}}  + r^2 {\rm d}\Omega_2 ^2 \quad\text{and}\quad B_{\mu \nu} = Q\frac{\epsilon_{\mu \nu}}{4\pi r^2}.
\end{equation}
Here, $\epsilon_{\mu \nu}$ is the volume form on the sphere; the $B$-field induces a vanishing NS-NS flux. Interestingly, since in 4D, the NS-NS flux can be traded for an axion pseudoscalar, Eq.~(\ref{BGHHS}) is the unique static Lorentzian axionic black hole solution with flat asymptotics.  One can analytically continue (\ref{BGHHS}) to make time Euclidean.
While the $B$-field in \cite{Bowick:1988xh} threads the two-sphere of the Schwarzschild black hole, one can construct a similar solution directly in the Euclidean signature in which the field has legs in the $\tau$ and $\phi$ dimensions: 
\begin{align}
\label{CigarAxion}
        {\rm d}s^2 = \left(1-\frac{2M}{r}\right){\rm d}\tau^2 + \frac{{\rm d}r^2}{1-\frac{2M}{r}}  + r^2 {\rm d}\Omega_2 ^2 \quad\text{and}\quad B_{\tau  \phi} = \frac{Q}{2\pi \beta}.
        \end{align}
Furthermore, it is possible to start with the Kerr Lorentzian solution, supplement it with the axion charge as these authors did, and analytically continue this geometry (both time $t\to -i\tau $ and the angular momentum parameter $a\to i\alpha$). This procedure results in the same $B$-field in Eq.~(\ref{CigarAxion}) and the following expression for the metric \cite{GibbonsHawking79}
\begin{subequations}
\begin{multline}
	{\rm d}s^2 = \frac{\Delta(r)}{r^2 -\alpha^2 \cos^2\theta} \left({\rm d}\tau + \alpha \sin^2\theta {\rm d}\phi\right)^2+(r^2 - \alpha^2 \cos^2 \theta)\left[\frac{{\rm d}r^2}{\Delta(r)} +{\rm d}\theta^2 \right]
 \\
 + \frac{\sin^2\theta}{r^2 -\alpha^2 \cos^2\theta}\left[(r^2 -\alpha^2){\rm d}\phi-\alpha {\rm d}\tau\right]^2\,,
\end{multline}
where
\begin{equation}
	\Delta(r) \equiv r^2 -2Mr - \alpha^2 ~,~
	r_+ \equiv M+\sqrt{M^2 + \alpha^2}.
\end{equation}
\end{subequations}%
To avoid a singularity at the horizon, the following identifications are required $(\tau,r,\theta,\phi)\sim (\tau+\beta,r,\theta,\phi+\beta \Omega)$ with
\begin{equation}
	\beta = \frac{4\pi M r_+}{\sqrt{M^2+\alpha^2}}\quad\text{and}\quad
	\Omega = \frac{\alpha}{r_+^2 - \alpha^2}.
\end{equation}
An apparent puzzle is that the $B$-field we consider is uniform in space, whereas in the high-temperature phase, it is a nontrivial function of $(r,\theta)$.  However, since we do not know to track the evolution of the system as a function of temperature or coupling (because for example, stringy effects become important at the correspondence point), this uniformization of the $B$-field is not ruled out by any physical law we are familiar with. We point out that the constant $B$-field and Ricci flatness of the low-temperature solution are consistent with the Green-Schwarz anomaly cancellation condition for the heterotic string\footnote{We would like to thank Emil Martinec for pointing out this condition.}.

In \cite{Roberto}, it was argued that a good control parameter to understand the transition between gravitationally bound strings and black holes is simply $g_s^2 S$. In particular, for large enough $g_s^2 S$, we expect the system to exhibit black hole behavior, whereas for small values of $g_s^2 S$, we expect to find a stringy behavior. We can test these ideas with our novel solutions. For the low-temperature phase, we find
\begin{equation}
\label{Thermodynamics}
\left(\frac{M}{M_p}\right)^2=\frac{\pi}{S} \left(\frac{S^2}{4 \pi ^2}-J^2\right)\,,
\end{equation}
whereas for the string phase, we have
\begin{equation}
\label{ThermodynamicsS}
\left(\frac{M}{M_p}\right)^2=\frac{3-2 \sqrt{2}}{\pi } g_s^2 S^2\,.
\end{equation}
For $J=0$, we can easily see that the two expressions will agree when $g_s^2 S\approx 1.457$, suggesting that there is indeed a transition between the two systems. Indeed, there is also no puzzle if we start in the black hole phase and reduce $g_s^2 S$, since there will be a value of $g_s^2 S$ at which $\left(\frac{M}{M_p}\right)^2$ is equal in both phases. There is, however, a puzzle if we start with a string configuration that carries large angular momentum. In particular, we note that for our rotating self-gravitating strings:
\begin{equation}
\frac{J}{S}=\frac{3 \sqrt{2}-4}{8 \pi } m_{\phi }\approx 0.00965\, m_{\phi}
\end{equation}
By taking $m_{\phi}>16$, one can consider rotating self-gravitating strings with $J> \frac{S}{2\pi}$, for which no corresponding black hole phase appears to exist. In this case, we can envisage three possibilities. 

In the first scenario, the fast rotating strings shed angular momentum to null infinity via short string emission \footnote{We would like to thank David~Turton for suggesting this possibility to us.}. This is perhaps a rather natural mechanism, especially in light of superradiance. Once the angular momentum is below $\frac{S}{2\pi}$ and one continues to increase $g_s^2 S$, one moves to the black hole phase. Since it involves radiating angular momentum to null infinity, this process appears non-adiabatic.

The second scenario is that one might be able to find hybrid-type configurations (see \cite{Frolov:1996xw,Snajdr:2002aa,Kinoshita:2016lqd,Igata:2018kry,Xing:2020ecz,Deng:2023cwh} for a description of such configurations), which are allowed to exist for larger values of $J$ at fixed entropy $S$, than the low-temperature phase we have considered here. This last scenario was first proposed in \cite{Roberto}.

The third possibility is that a transition to the black hole does not take place; in conjunction with the condition $J>\frac{S}{2\pi}$, Eq~(\ref{Bscaling}) implies that the $B$-field increases when decreasing the temperature, which generates a repulsive force that might counterbalance the gravitational attraction to the center of mass. 

To summarize this subsection, we have characterized an analytical continuation of an axionic Kerr black hole solution that can be connected to the self-gravitating string condensate as the temperature is varied.

\section{Conclusions}
\label{Sec:Conc}

We have constructed a rotating version of the $D=4$ Horowitz-Polchinski self-gravitating string in heterotic string theory. The solution is weakly-coupled, weakly-curved, and valid near the Hagedorn temperature. The winding-momentum condensate depends on both the radial and angular coordinates, rotating within a plane parameterized by the azimuthal angle relative to Euclidean time. We have obtained a discrete family of solutions parametrised by an integer $m_{\phi}$, which determines the phase of the condensate. As $m_{\phi}$ increases, the string condensate becomes concentrated further away from the rotation axis due to the centrifugal force, as expected.

We calculated the entropy, angular momentum, and NS-NS charge of the solution at tree-level, and found that, to leading order in the deviation from the Hagedorn temperature, these are all proportional to the energy of the string condensate.  

When the temperature is significantly lower than the Hagedorn temperature, a Euclidean solution sharing the same charges corresponds to the doubly analytically continued Kerr black hole, augmented with axion charge. The double continuation is defined via $t\to -i \tau~,~a\to i \alpha$,  and when applied for the conventional Kerr solution, admits an interpretation in terms of a Kaluza-Klein monopole-anti-monopole pair \cite{GrossPerry83}. The authors of \cite{Dowker95} provided a Lorentzian interpretation of a similar Euclidean configuration, describing the pair creation of black holes in a magnetic field. In our context, we imagine a (Euclidean signature) dyonic monopole that couples to the graviphoton and the $B$-field, which has a Lorentzian interpretation describing pair creation of heterotic strings.  

We endeavored to construct self-gravitating rotating strings in Type II string theory, but our attempts proved unsuccessful. Currently, we posit that such a construction is unattainable using the leading order action of Type II, though  $\alpha^\prime$ corrections could change this conclusion. The absence of obstruction in heterotic strings stems from the unique property of the thermal scalar $\chi$, which exhibits charge under \emph{both} the $B$ field and the graviphoton $A$, unlike in Type II. Intriguingly, heterotic strings also play a special role in \cite{CMW}, prompting an exploration into whether this connection extends beyond mere coincidence.

We interpret our solution as describing strings near the Hagedorn temperature, which experience a Newtonian gravitational potential and are charged with respect to the $B$-field. The Lorentzian interpretation has mysterious aspects since an imaginary angular momentum and an imaginary $B$-field emerge. It would be interesting to find out how to construct a Horowitz-Polchinski self-gravitating string that has nonzero and \emph{purely imaginary} angular momentum in Euclidean signature, which continues to a real angular momentum in Lorentzian signature.

\section*{Acknowledgements}
We thank Ramy~Brustein, Yiming~Chen, Gary~Horowitz, Roberto~Emparan, Raghu~Mahajan, Emil~Martinec, Kostas~Skenderis, Marija~Tomašević and David~Turton for helpful discussions, and \'Oscar~Dias for comments on an earlier version of this work. The work of J.~E.~Santos is partially supported by STFC consolidated grants ST/T000694/1 and ST/X000664/1. Y.~Z. is supported by the Blavatnik Postdoctoral Fellowship.

\appendix

\section{ Covariant Derivative Formula}
\label{App:A}

The purpose of this appendix is to perform a string S-matrix computation in the heterotic theory that leads to three-point couplings in the effective action that appear in $D_{\mu} \chi D^{\mu} \chi^*$, with
\begin{equation}
\label{CovariantDerivative2}
	D_{\mu} \chi = \partial_{\mu} \chi + i \frac{\beta}{2\pi \alpha'}B_{\tau \mu}\chi +i\frac{\pi }{\beta}A_{\mu} \chi.
\end{equation}
 The importance of this form of the covariant derivative is that the $i$ factors imply that the solution we find in this paper has real angular momentum in Euclidean signature. As shown below, the $i$ factors in Eq.~(\ref{CovariantDerivative2}) follow from the conventional transition between momenta vectors that appear in the string amplitudes to $~-i\times$ partial spatial derivatives on fields in the effective action. The result we obtain is identical to \cite{SchulginTroost}, though as we will explain, the calculation we present is slightly different. We adopt the conventions of Polchinski's books \cite{Polchinski1}, \cite{Polchinski2}.

To achieve the desired three-point couplings, we calculate the amplitude of two winding-momentum modes with opposite quantum numbers $w=\pm 1~,~n=\pm \frac{1}{2}$ to produce a massless mode, which can be either a graviton, radion, dilaton,  Kalb-Ramond or graviphoton. We will be interested in the latter two modes, that give rise to the three-point couplings in the squared of the covariant derivatives $D_{\mu} \chi D^{\mu} \chi^*$. 

The radius of the thermal circle is related to its circumference, 
\begin{equation}
    R=\frac{\beta}{2\pi},
\end{equation}
and is described by the field 
\begin{equation}
\label{Decomposition}
X^0(z,\bar{z}) =X^0 _L (z)+X^0 _R (\bar{z}),    
\end{equation}
which is written in Eq.~(\ref{Decomposition}) as a sum of the holomorphic part and the antiholomorphic part. Periodic boundary conditions are imposed for target-space bosons and anti-periodic boundary conditions for target-space fermions. 

In the heterotic string on a planer worldsheet geometry, local vertex operators creating the $w=1,n=\frac{1}{2}$ winding-momentum mode with respect to the thermal circle can be written in the $(-1)$ picture as
\begin{equation}
 \label{w1n1/2}
    V^{ 1, 1/2} (z,\bar{z})= g_c 'e^{i\vec{k}_{\perp} \cdot \vec{X}(z,\bar{z})} e^{i k_L  X^0 _L (z)+i k_R X^0 _R (\bar{z}) }e^{-\tilde{\phi}(\bar{z})} ~,~ 
\end{equation}
\begin{equation}
 k_{L} = \frac{ R}{\alpha'} + \frac{1}{2R} ~,~  k_{R} = -\frac{ R}{\alpha'} + \frac{1}{2R}. 
\end{equation}
The notations in Eq.~(\ref{w1n1/2}) include $\vec{k}_{\perp}$, a spatial momentum of the string state, the superscript $\perp$ denotes ``transverse to the circle'', and $e^{-\tilde{\phi} (\tilde{z})}$ represents a ghost associated with fixing an anti-commuting coordinate. The coupling $g_c ' $ is defined in terms of the gravitational Newton constant $G_N ^{(D)}$ and $\beta$ as follows:
\begin{equation}
  g_c ' = \frac{\kappa'}{2\pi} ~,~ \kappa' \equiv \frac{\kappa}{\sqrt{\beta}} ~,~ \kappa^2 \equiv 8\pi G_N.
\end{equation}
For $w=-1$ and $n=-\frac{1}{2}$, one has instead
\begin{equation}
    V^{ -1, -1/2} (z,\bar{z})= g_c ' e^{i\vec{k}_{\perp} \cdot \vec{X}(z,\bar{z})} e^{i k_L  X^0 _L (z,\bar{z})+i k_R X^0 _R (z,\bar{z}) }e^{-\tilde{\phi}(\bar{z})} ~,~ 
\end{equation}
\begin{equation}
    k_{L} = - \frac{ R}{\alpha'} - \frac{1}{2R} ~,~ k_R =   \frac{ R}{\alpha'} - \frac{1}{2R}.
\end{equation}
Closed massless heterotic modes can be created by acting with the following vertex operator, with the polarization tensor components $e_{\mu \nu}$ and momentum $\vec{k}$,  written in the $0$ picture:
\begin{equation}
 V^{(0)} (e,k,z,\bar{z}) = \frac{2 g_c '}{\alpha'} e_{\mu \nu}  \partial X^{\mu} (z) \left( \bar{\partial} X^{\nu} -\frac{1}{2}i \alpha' \vec{k}\cdot \tilde{\psi} \tilde{\psi} ^{\nu} (\bar{z})\right)e^{i \vec{k}\cdot \vec{X}(z,\bar{z})}. 
\end{equation}
The right-moving worldsheet fermionic field $\tilde{\psi}(\bar{z})$ that appears in the last equation will not play a role because it will not have other fermion partners to contract with. While reference \cite{SchulginTroost} computed the same amplitude we want to compute, they chose to distribute the pictures as $\{-1,0\}$ for the winding-momentum modes, and $-1$ for the massless modes; we choose instead the $\{-1,-1\}$ and $0$ pictures respectively for these string states.

We require the two-point function of the ghost operators
\begin{equation}
\label{Ghosts2Pt}
  \left \langle e^{-\tilde{\phi}(\bar{z}_1)} e^{-\tilde{\phi}(\bar{z}_2)} \right \rangle = \frac{1}{\bar{z}_{12}}.
\end{equation}
Also, the three-point function of the ghosts which allow one to fix the worldsheet positions of the vertex operators is
\begin{equation}
\label{Part1}
    \langle c(z_1)\tilde{c}(\bar{z}_1) c(z_2) \tilde{c} (\bar{z}_2)  c(z_3) \tilde{c} (\bar{z}_3)\rangle =|z_{12}|^2 |z_{23}|^2 |z_{13}|^2.
\end{equation}
The string amplitude is 
\begin{equation}
\label{Amplitude}
    S(k_1,k_2,k_3,e_3) = \left\langle [V^{1,1/2}(z_1,\bar{z}_1)]_r [V^{-1,-1/2}(z_2,\bar{z}_2)]_r [V^{(0)} (z_3,\bar{z}_3)]_r c\tilde{c}(z_1,\bar{z}_1) c\tilde{c}(z_2,\bar{z}_2) c\tilde{c}(z_3,\bar{z}_3) \right\rangle.
\end{equation}
The notation $[]_r$ means ``renormalized'', removing UV divergences from internal contractions within the composite vertex operators. 
The result includes a contribution from zero modes of the $X$-theory that enforces momentum conservation transverse to the circle, times the volume of this $S^1 _{\beta}$:
\begin{equation}
\label{Part2}
 \text{Zero-Modes} = \beta (2\pi)^{d} \delta^d (k_1 ^{\perp} + k_2 ^{\perp} + k_3 ^{\perp}).
\end{equation}
One can incorporate a $T^{9-d}$ compactification manifold by multiplying the latter equation by the volume of this torus.
The nonzero modes in the $X$-theory give rise to the following factor in the correlation function, with a prime indicating that zero modes are not included:
\begin{align}
&(g_c ')^2 \frac{2 g_c '}{\alpha'}\left \langle [e^{i\vec{k}^{\perp} _1 \cdot \vec{X}(z_1,\bar{z}_1)} e^{i k_{1L}  X^0 _L (z_1)+i k_{1R} X^0 _{1R} (\bar{z}_1) }]_r [e^{i\vec{k}^{\perp} _2 \cdot \vec{X}(z_2,\bar{z}_2)} e^{i k_{2L}  X^0 _L (z_2)+i k_{2R} X^0 _{2R} (\bar{z}_2) }]_r  \right. \nonumber\\
& \left. e_{\mu \nu}  [\partial X^{\mu}(z_3,\bar{z}_3)  \bar{\partial} X^{\nu} (z_3,\bar{z}_3) e^{i \vec{k}_3 ^{\perp} \cdot X(z_3,\bar{z}_3)}]_r \right\rangle ' =e_{\mu \nu}~\frac{2(g_c')^3}{\alpha'}\times \frac{8\pi}{(g_c ')^2 \alpha'}\times \left(-i\frac{\alpha'}{2}\right)^2 \nonumber\\
&\times \left[\frac{k_1 ^{\mu}}{z_{13}} +\frac{k_2 ^{\mu}}{z_{23}} \right]\left[\frac{k_1 ^{\nu}}{\bar{z}_{13}}  +\frac{k_2 ^{\nu}}{\bar{z}_{23}} \right]|{z}_{13}| ^{\alpha' k_1 ^{\perp}\cdot k_3 ^{\perp} } |{z}_{23}| ^{\alpha' k_2 ^{\perp}\cdot k_3 ^{\perp} }z_{12} ^{\frac{\alpha'}{2}k_{1L}\cdot k_{2L}}\bar{z}_{12} ^{\frac{\alpha'}{2}k_{1R}\cdot k_{2R}} .
\end{align}
In the second line, the factor of $\frac{8\pi}{(g_c ')^2 \alpha'}$ arises from an overall normalization of sphere correlation functions \cite{Polchinski1}. Next, one can utilise the fact that the polarization tensor $e_{\mu\nu}$ is transverse to the momentum vector $k_3 ^{\mu}$ and exploit momentum conservation. These imply that $e_{\mu \nu} k_1 ^{\mu} = \frac{1}{2}e_{\mu \nu} (k_1 ^{\mu} - k_2 ^{\mu})$ and $e_{\mu \nu} k_2 ^{\mu} = -\frac{1}{2}e_{\mu \nu} (k_1 ^{\mu} - k_2 ^{\mu})$. Furthermore, the on-shell condition imposes $k_3 ^2=0$. As a result,
\begin{align}
&\left\langle [V^{1,1/2}(z_1,\bar{z}_1)]_r [V^{-1,-1/2}(z_2,\bar{z}_2)]_r [V^{(0)} (z_3,\bar{z}_3)]_r \right\rangle'=\nonumber\\
&\frac{1}{\bar{z}_{12}}(-\pi g_c ') e_{\mu \nu} (k_1 ^{\mu} - k_2 ^{\mu})(k_1 ^{\nu} - k_2 ^{\nu}) \frac{|z_{12}|^2}{|z_{13}|^2 |z_{23}|^2} z_{12} ^{-\frac{\alpha'}{2}\left(\frac{R}{\alpha'} +\frac{1}{2R} \right)^2} \bar{z}_{12} ^{-\frac{\alpha'}{2}\left(\frac{R}{\alpha'} -\frac{1}{2R} \right)^2} .
\end{align}
The first $\frac{1}{\bar{z}_{12}}$ comes from the two-point function of the ghosts $e^{-\tilde{\phi}}$, see Eq.~(\ref{Ghosts2Pt}).
The winding-momentum modes in this system are on-shell at the Hagedorn temperature where $m^2=0$. This ties the radius $R$ with $\alpha'$ \cite{HeteroticStringFree},\cite{AW}:
\begin{equation}
    \frac{R^2}{(\alpha')^2} +\frac{1}{4R^2}=\frac{3}{\alpha'}.
\end{equation}
Therefore,
\begin{align}
\label{Part3}
&\left\langle [V^{1,1/2}(z_1,\bar{z}_1)]_r [V^{-1,-1/2}(z_2,\bar{z}_2)]_r [V^{(0)} (z_3,\bar{z}_3)]_r \right\rangle'=-\pi g_c ' e_{\mu \nu}  \frac{(k_1 ^{\mu} - k_2 ^{\mu})(k_1 ^{\nu} - k_2 ^{\nu})}{|z_{12}|^2|z_{13}|^2 |z_{23}|^2}.
\end{align}
Combining Eqs. (\ref{Part1}),(\ref{Part2}) and (\ref{Part3}) in Eq.~(\ref{Amplitude}) results in:
\begin{equation}
 S(k_1,k_2,k_3,e_3) = -2\kappa' e_{\mu \nu} k_1 ^{\mu} k_2 ^{\nu}  \beta (2\pi)^d \delta^{d} \left(k_1 ^{\perp}+k_2 ^{\perp} + k_3 ^{\perp} \right). 
\end{equation}
This same result can also be obtained for for the bosonic and Type II string theories \cite{RamyYoav3}.

We want to extract the corresponding vertex in the effective field theory. The Euclidean action appears in the path integral as $e^{-S_{vertex}}$ so that the amplitude contains a minus sign from expanding the exponential in Taylor series.
Second, the factor of 
\begin{equation}
(2\pi)^d\delta^d \left( k_1 ^{\perp}+k_2 ^{\perp}+k_3 ^{\perp}\right)\to \int {\rm d}^d x.
\end{equation}
Third, the factor of $(-2\kappa')$ stays intact.  We pick the polarization tensor 
\begin{equation}
\frac{1}{\sqrt{2}}\left(e_{\tau \mu} - e_{\mu \tau}\right),	
\end{equation}
where $\mu$ is a spatial index - this corresponds to the $B_{\tau \mu}$ field. 
The momenta $k_{1,2} ^{\mu}$ transform to
\begin{equation}
	k_{1,2~\perp} ^{\mu} \to -i \partial ^{\mu}.
\end{equation}
Now, the antisymmetric polarization tensor picks up the winding part of the 0'th component of $k$: $\frac{R}{\alpha'}=\frac{\beta}{2\pi \alpha'} $.
This procedure gives the amplitude for a target space action with a normalized kinetic term for the $B$-field. However, typically one further performs 
\begin{equation}
B_{\tau \mu}^{normalized}\to B_{\tau \mu}^{Polchinski}=\frac{1}{\sqrt{2} \kappa'} B_{\tau \mu} ^{normalized},	
\end{equation}
in order to match the normalization of this field in Polchinski's conventions (\ref{I_2}).

Following these steps, one obtains the desired three-point coupling
\begin{equation}
I_{B\chi \chi^*}=\beta \frac{i\beta}{2\pi \alpha'}\int {\rm d}^d x  B_{\tau \mu} \left( \chi \partial^{\mu} \chi ^* - \chi^* \partial^{\mu} \chi\right).
\end{equation}
Similarly, the three-point coupling involving the graviphoton is
\begin{equation}
I_{A\chi \chi^*}=\beta \frac{i\pi}{\beta}\int {\rm d}^d x  A_{ \mu} \left( \chi \partial^{\mu} \chi ^* - \chi^* \partial^{\mu} \chi\right).
\end{equation}

\section{Alternative Writing of the Equations}
We present the equations of motion in a form where no dimensional reduction with respect to $S^1 _{\beta}$ is done for the supergravity sector.
The light fields are denoted by 
\begin{align}
\label{Fields2}
 \{\Phi_D,G_{IJ}, B_{IJ}, \chi,\chi^* \} ~,~ I,J=1,...,d+1,
\end{align} 
where $\Phi_D$ is the $D$-dimensional dilaton, $G_{IJ}$ are metric components in the string frame, $B_{IJ}$ are NS-NS two-form components. 
The two-term appearing in the action describing the system are
	 \begin{align}
  \label{I_1app}
	I_1 = \beta \int {\rm d}^d x \sqrt{G_D} e^{-2\Phi_D}\left[ D_{\mu} \chi D^{\mu} \chi^* +\left( \frac{\beta ^2 G_{\tau \tau} }{4\pi^2 {\alpha^\prime}^2}+G^{\tau \tau}\frac{\pi^2 }{\beta^2\,}-\frac{3}{\alpha^\prime}\right) \chi \chi^*\right]\,,
\end{align}
the covariant derivative $D_{\mu} \chi$ is defined below in Eq.~(\ref{CovariantDerivative}), and $G_{\tau \tau}$ is the metric component with two legs in the Euclidean time direction.

The NS-NS sector of supergravity fields is described by:
\begin{equation}
	\label{I_2app}
	I_2 = -\frac{\beta}{2\kappa_0 ^2}\int {\rm d}^d x \sqrt{G_D} e^{-2\Phi_D} \left[ R_D + 4\partial_{I} \Phi_D \partial^{I} \Phi_D -\frac{1}{12} H_{IJK} H^{IJK}\right].
\end{equation}
Here, $I,J,K=1,...,D$ are indices that run over the spatial non-compact dimensions and Euclidean time. The normalization factor $\frac{1}{\kappa_0 ^2}$ combines with the asymptotic value of the dilaton to produce the inverse of the gravitational constant: $\frac{1}{8\pi G_N} = \frac{1}{\kappa_0 ^2 e^{2\Phi(\infty)}}$.

The equations of motion are written below. 
\begin{itemize}
	\item
	{\em The $\chi$ equation}
	\begin{eqnarray}
		\label{chiEOMhetapp}
		&\frac{e^{2\Phi_D}}{\sqrt{G_D}}\partial_{\mu}\left(e^{-2\Phi_D}\sqrt{G_D}G^{\mu \nu} D_{\nu}\chi \right)+i\left( \frac{\beta}{2\pi \alpha'} B_{\tau}^{~ \mu} + \frac{\pi}{\beta} A^{\mu}\right)D_{\mu} \chi =\left[\frac{\beta ^2 G_{\tau \tau} }{4\pi^2 (\alpha')^2}+\frac{\pi^2}{\beta^2} G^{\tau \tau}-\frac{3}{\alpha^\prime} \right]\chi.\nonumber\\
  &
	\end{eqnarray}	
	\item
	{\em The Kalb-Ramond equations}
\begin{equation}
	\label{KalbRamondapp}
\frac{e^{2\Phi_D}}{\sqrt{G_D}} \partial_I \left( \sqrt{G_D} e^{-2\Phi_D}H^{IJK}\right)=\delta_{J,\tau}\frac{i\beta \kappa_0 ^2}{\pi \alpha'}  \left( \chi D_K \chi^* - \chi^* D_K \chi\right). 
\end{equation}
	\item
{\em The dilaton equation}
\begin{align}
\label{dilatonapp}
	&R_D + 4\nabla^2 \Phi_D-4\partial_{\mu}\Phi_D \partial^{\mu} \Phi_D - \frac{1}{12} H_{IJK} H^{IJK} =\nonumber\\
	&2\kappa_0 ^2  \left[ D ^{\mu} \chi D_{\mu} \chi^* +\left(\frac{\beta ^2 G_{\tau \tau}}{4\pi^2 (\alpha')^2}+\frac{\pi^2}{\beta^2 }G^{\tau \tau} -\frac{3}{\alpha^\prime} \right)\chi \chi^*\right].
\end{align}
\item
{\em A linear combination of the spatial metric and dilaton equations}
\begin{eqnarray}
	\label{EinsteinRhet2app}
	&(R_D)_{IJ} +2\nabla_{I}  \nabla_{J} \Phi_D -\frac{1}{4}H_{ IKL}H_{J}^{~~ KL}   =2\kappa_0 ^2 D_{I} \chi D_{J} \chi^*.
\end{eqnarray}
\end{itemize}

\section{Attempt to Find a Smooth Rotating Type II Solution}
When attempting to find a smooth solution in Type II, we encountered the following issues. In this theory, the winding mode that is light near the Hagedorn temperature does not carry momentum, therefore the graviphoton equation of motion derived from the conventional HP action \cite{HP97} can be written as
\begin{equation}
 \partial _{\mu}  \left( e^{-2\Phi_d+2\sigma} \sqrt{G_d} F^{\mu \nu}\right)  =0.
\end{equation}
We now set $d=3$. A simple solution to this equation has the expression in the wiggly brackets constant. In the HP solution \cite{HP97}, this constant vanishes - corresponding to a zero angular momentum configuration. When the constant does not vanish, and in the approximations of small dilaton and radion, one can consider
\begin{equation}
\label{FII}
 F^{r\phi} \approx \frac{\text{const}}{r^2}.
\end{equation}
This is problematic since $F_{r \phi} F^{\phi r}\propto \frac{1}{r^2}$ diverges at the origin. There is no analogous problem for the Euclidean black holes in which one considers radial coordinates bigger than the horizon radius. An attempt to circumvent the problem is to apply an NS-NS field strength $H_{\tau \mu \nu} \neq0$, for instance by turning on a $B_{\tau \theta}$ component. This generates a nonzero right-hand-side:
\begin{eqnarray}
 \label{GraviphotonII}
	&\frac{e^{2\Phi_d}}{\sqrt{G_d}}\partial_{\mu} \left( \sqrt{G_d} e^{-2\Phi_d+2\sigma} F^{\mu \nu}\right) =-\frac{1}{2} \tilde{H}^{\nu \lambda \mu} H_{\tau\lambda \mu},
	\end{eqnarray}
 where $\tilde{H}$ is defined in Eq.~(\ref{TildeH}).
However, the application of $H_{\tau \mu \nu}$ gives an effective mass term to $A_{\phi}$, obstructing the asymptotic power-law behavior that characterizes rotating metrics. Another idea is adding a momentum condensate with momentum number $n=2$ (and zero oscillator numbers) to resolve this singularity. However, this mode does not exist in Type II with thermal boundary conditions. \footnote{These obstructions might be akin to the surprise expressed by \cite{CMW}, that there is a difference between Type II superstring theory and the heterotic string theory.}  Possibly, the singularity that follows from Eq.~(\ref{FII}) can be resolved through $\alpha'$ corrections, similarly to \cite{Sen24},\cite{YimingNew}.  

\section{Numerical Convergence\label{app:conv}}
\begin{figure}[t]
    \centering\includegraphics[height=6cm]{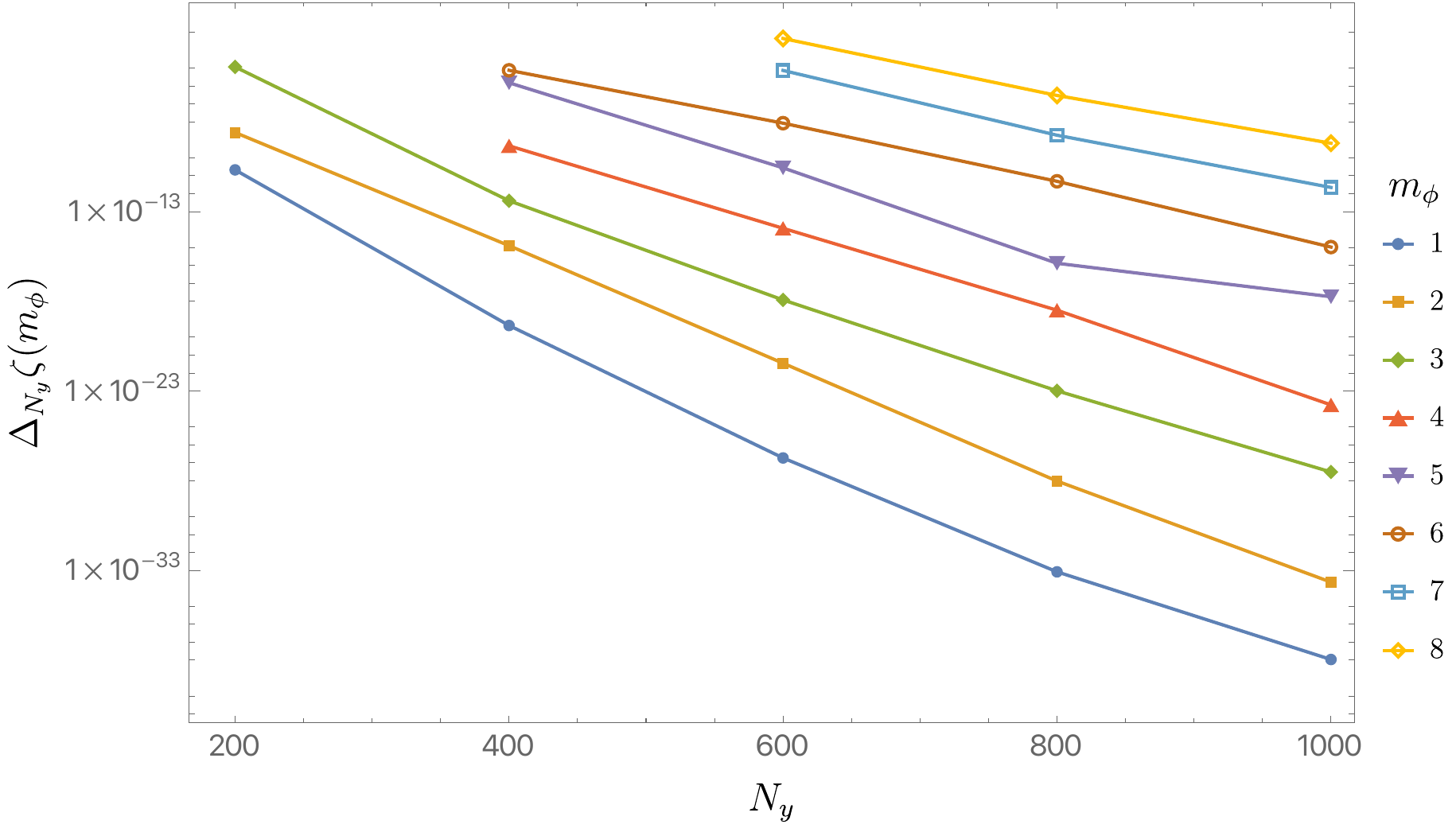}
    \caption{$\Delta_{N_y} \zeta(m_{\phi})$ in a logarithmic scale, as a function of $N_y$ for different values of $m_{\phi}$ for  fixed $N_{\theta}=12$.}
    \label{fig:con}
\end{figure}

In this section, we study how numerical results depend on the discretisation scheme. For each value of $m_{\phi}$, there are two dials we can change to study the convergence of our numerical method: $N_{\theta}$ (the number of terms in our expansion in Legendre and associated Legendre polynomials) and $N_y$ (the number of points in the $y$ direction). We will start by analysing the convergence in $N_y$, by defining the quantity
\begin{equation}
\Delta_{N_y} \zeta(m_{\phi})\equiv 100 \left|1-\frac{\zeta(m_{\phi})_{N_y}}{\zeta(m_{\phi})_{N_y+200}}\right|
\end{equation}
where $\zeta(m_{\phi})_{N_y}$ is computed on a collocation grid with $N_y$ points at fixed $N_{\theta}$. In Fig.~\ref{fig:con} we plot $\Delta_{N_y} \zeta(m_{\phi})$ as a function of $N_y$ in a logarithmic scale for fixed $N_{\theta}=10$. The trend indicates exponential convergence in $N_y$. We note that for increasingly large values of $m_{\phi}$, we can only find solutions for large enough $N_y$. The reason for this is that the solutions localise near $y=1$, and as such need increasing resolution to resolve these.

\begin{figure}[t]
    \centering\includegraphics[height=6cm]{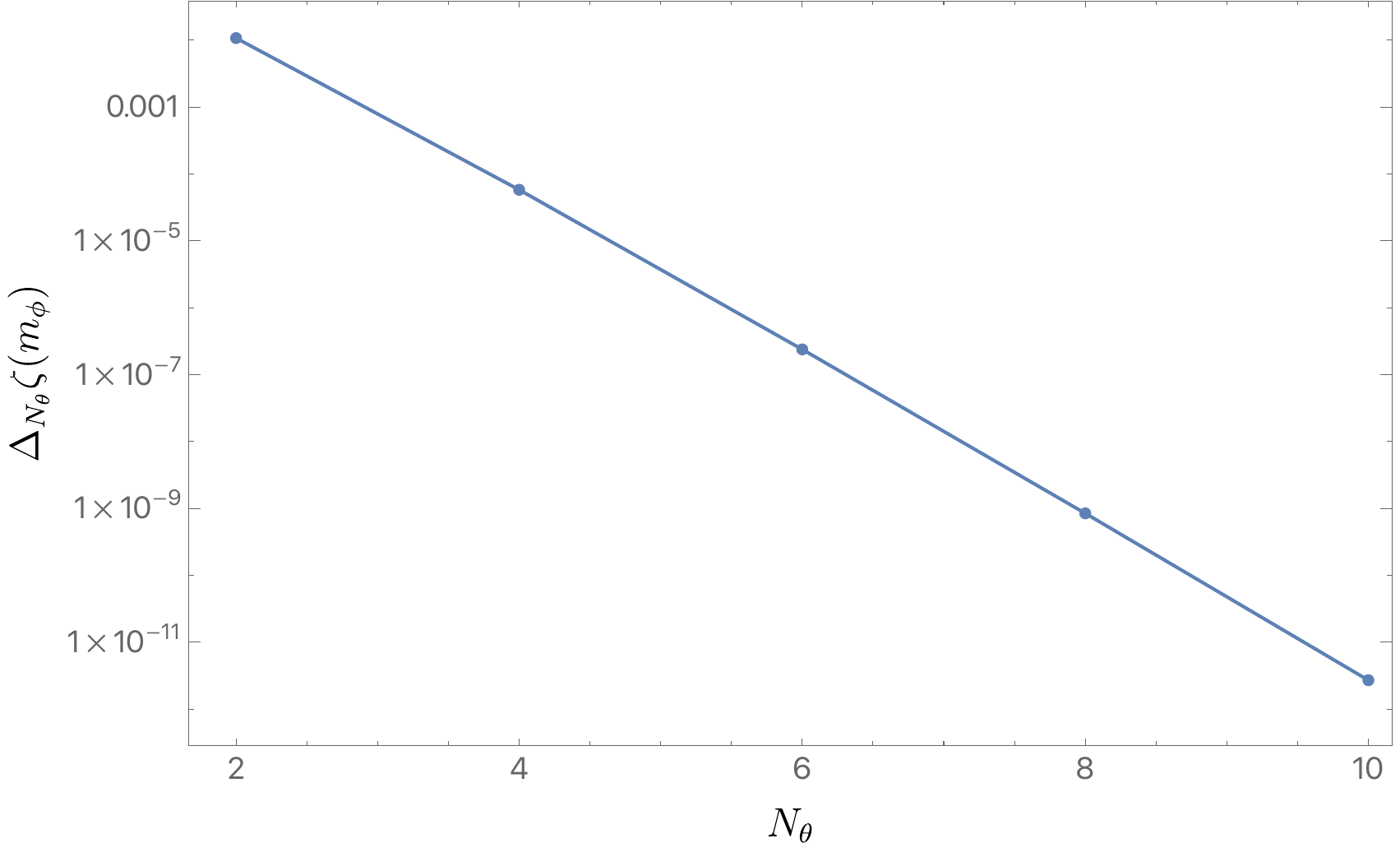}
    \caption{$\Delta_{N_{\theta}} \zeta(m_{\phi})$ in a logarithmic scale, as a function of $N_{\theta}$ for fix $m_{\phi}=1$ and $N_{y}=200$.}
    \label{fig:Ntheta}
\end{figure}

Finally, we turn our attention to changing $N_{\theta}$. As our WKB analysis shows, for large enough $m_{\phi}$ we expect the solutions to localised around a particular associated Legendre polynomial. As such, when $m_{\phi}$ is large enough $N_y$ is the most important dial, which we studied extensively in Fig.~\ref{fig:con}. However, for smaller values of $m_{\phi}$ it is worth exploring the dependence in $N_{\theta}$. In Fig.~\ref{fig:Ntheta} we plot
\begin{equation}
\Delta_{N_{\theta}} \zeta(m_{\phi})\equiv 100 \left|1-\frac{\zeta(m_{\phi})_{N_{\theta}}}{\zeta(m_{\phi})_{N_{\theta}+2}}\right|
\end{equation}
in a logarithmic scale for $m_{\phi}=1$ and for fixed $N_{y}=200$. Again, we clearly see evidence for exponential convergence in $N_{\theta}$.

\newpage 

\bibliography{hp}{}
\bibliographystyle{JHEP}

\end{document}